\documentclass[preprint,12pt]{elsarticle}

\usepackage{epsfig}
\usepackage{imakeidx}
\makeindex
\usepackage[T1]{fontenc} 

\usepackage{graphicx}
\usepackage[margin=1in]{geometry}
\usepackage[utf8]{inputenc}
\usepackage[section]{placeins}
\usepackage{color}
\usepackage{hyperref}
\usepackage{eqnarray}
\usepackage{tikz}
\usepackage{subfig}
\usepackage{tensor}
\usepackage{amsmath,amsthm,amssymb}
\usepackage[ruled,vlined]{algorithm2e}
\usepackage{multicol}
\usepackage{float}
\usepackage{blindtext}
\usepackage{makecell}

\journal{Annals of Physics}

\begin{document}

\begin{frontmatter}



\title{Observational Constraints on Chaplygin Gas Models in Non-Minimally Coupled Power Law \texorpdfstring{$f(Q)$}{f(Q)} Gravity with Quasars}
\author[a]{Nakul Aggarwal \corref{1}}
\ead{na935@physics.rutgers.edu}
\author[a]{Ali Pourmand \corref{1}}
\ead{pourmand@ualberta.ca}

\author[b]{Fatimah Shojai \corref{2}}
\ead{fshojai@ut.ac.ir}

\author[c]{Harish Parthasarathy}
\ead{harishp@nsit.ac.in}

\cortext[1]{Equal contribution}
\cortext[2]{Corresponding Author}
\affiliation[a]{organization={Department of Physics, University of Alberta}, city={Edmonton}, postcode={T6G 2E7}, state={Alberta}, country={Canada}}
\affiliation[b]{organization={Department of Physics, University of Tehran}, postcode={P.O. Box 14395-547},city={Tehran},country={Iran}}
\affiliation[c]{organization={Department of Electronics and Communication Engineering, Netaji Subhas University of Technology}, postcode={110078}, city={New Delhi},country={India}}


\begin{abstract}
In the framework of \texorpdfstring{$f(Q)$}{fq} gravity, where gravity emerges from non-metricity \texorpdfstring{$Q$}{q}, we explore the cosmological implications of its non-minimal coupling to matter. Inspired by the recent success of Chaplygin gas models in explaining dark energy, we consider a background fluid composed of baryonic matter, radiation, and a family of Chaplygin gas variants namely Generalized Chaplygin Gas (GCG), Modified Chaplygin Gas (MCG), and Variable Chaplygin Gas (VCG). We constrain these models with three recent observational datasets: Observational Hubble Data (OHD), Baryonic Acoustic Oscillation (BAO) measurements, and Quasi-Stellar Objects (QSO) data. For the QSO dataset, we propose an analytical expression for errors in comoving distance to circumvent the reliance on Monte Carlo simulations. Using kinematic diagnostics such as the deceleration and jerk parameters and Om diagnostic, we assess deviations of the proposed models from \texorpdfstring{$\Lambda$CDM}{lcdm}. Our joint analysis of the three datasets reveals that the transition redshift from a decelerated to an accelerated expansion of the universe for the GCG, MCG and VCG models is \texorpdfstring{$0.620^{+0.018}_{-0.017}$}{1}, \texorpdfstring{$0.537^{+0.017}_{-0.017}$}{2} and \texorpdfstring{$0.470^{+0.012}_{-0.012}$}{3} respectively, indicating a departure from \texorpdfstring{$\Lambda$CDM}{lcdm}.
\end{abstract}


\end{frontmatter}

\section{Introduction}
\label{sec:intro}
Based on observations \cite{riess1998observational,perlmutter1999measurements,eisenstein2005detection,percival2010baryon}, it is a well-established fact that the universe is accelerating. One of the most plausible theories for this acceleration is the existence of dark energy with negative pressure. Of the wide range of candidates for dark energy, $\Lambda$CDM has gained prominence as a notably successful model. It has demonstrated the ability to explain several cosmological phenomena such as the formation of large-scale structures and provided accurate descriptions of the type Ia Supernovae (SNe Ia) observations. However, it suffers from major problems, namely fine-tuning and cosmic coincidence \cite{copeland2006dynamics}. 

To alleviate these problems, several candidates for dark energy have been proposed in addition to the cosmological constant, where the effective equation of state (EOS) of dark energy is a function of cosmic time, such as quintessence \cite{PhysRevD.37.3406, caldwell1998cosmological}, k-essence \cite{PhysRevD.62.023511}, Holographic Dark Energy (HDE) \cite{Li_2004}, and Chaplygin gas (CG) \cite{kamenshchik2001alternative}. The HDE model is theoretically attractive, for example, because it is based on the holographic principle \cite{hooft2009dimensionalreductionquantumgravity}, which is believed to be a fundamental principle of quantum gravity \cite{wang2017holographic} (for more information about the theoretical aspects of HDE, see \cite{Nojiri_2006,2021Astashenok}). This theory has several shortcomings however, such as the fact that there is a physical quantity in the model which is known as the characteristic length scale, which must be specified by making some assumptions (it is usually chosen to be equal to the future event horizon \cite{Li_2004}). Moreover, it is essentially only considering dark energy whereas other theoretical models have been proposed as attempts to solve the dark energy problem.\\
Alternatively, the CG model has been extensively explored as a compelling candidate for dark energy, offering an intriguing substitute to the standard perfect-fluid description of the universe. Its corresponding EOS, $p=-C/\rho$, where $p$ and $\rho$ are pressure and energy density, respectively, and $C$ is a positive constant, endows it with a dual nature: it mimics pressureless dark matter in the early universe and cosmological constant at late times. The CG model's strength is partly due to the fact that it provides both matter and dark energy through a single EOS, which makes it a simpler model compared to HDE because the Friedmann equations can be solved. Moreover, in \cite{xu2016comparison}, using the JLA compilation of Type Ia Supernovae alongside the Planck 2015 CMB measurements, Baryonic Acoustic Oscillations (BAO), and Observational Hubble Data (OHD) datasets, the authors demonstrated that the
Generalized Chaplygin Gas (GCG) model is statistically favored over the HDE framework, as reflected in the markedly lower value of its information criteria namely Akaike Information Criterion (AIC) and Bayesian Information Criterion (BIC) when compared to $\Lambda$CDM. In \cite{wu2025comparison} using a combination of Cosmic Chronometer measurements, BAO, Type Ia Supernovae, and Strong Gravitational Lensing Time-Delay observations, it was also found that the HDE model received comparatively weaker statistical support when compared to $\Lambda$CDM. Since at this stage there is currently no model that is both theoretically solid and consistent with all observational data, examining alternate models such as CG with its aforementioned advantages is warranted.

Despite its elegant simplicity, the CG model faces significant challenges in reproducing the observed cosmological power spectrum \cite{sandvik2004end}.
To address these shortcomings, the authors in \cite{kamenshchik2001alternative} introduced a phenomenological extension of CG that interpolates between the dust and dark energy dominated eras, namely the "Generalized Chaplygin Gas" (GCG) \cite{bento2002generalized}. GCG is a perfect fluid with a polytropic EOS $p=-C/\rho^n$, where $0<n\le1$ and $C$ is a positive quantity. However, it has also been shown to lead to instabilities in the perturbation spectrum, rendering it incompatible with large-scale structure formation \cite{sandvik2004end,amendola2003wmap}. 

To mitigate these instabilities, "Modified Chaplygin gas" (MCG) with its EOS $p=B\rho-C/\rho^n$ was proposed in \cite{benaoum2022accelerated}. It maintains a constant negative pressure at low-density and a high pressure in the high-density regime. The inclusion of the additional term in MCG suppresses unphysical oscillations \cite{e2008mathematical}. 
Another notable refinement, proposed in \cite{guo2007cosmology}, is the "Variable Chaplygin Gas" (VCG), wherein the parameter
$C$ evolves dynamically across cosmic epochs. This model was shown to align with the Born-Infeld tachyon action, establishing a framework in which dark matter and dark energy interact with each other. Using the gold sample of Type Ia supernova data and X-ray gas mass fraction measurements, the same authors in \cite{guo2005observational} ruled out CG at the confidence level $3\sigma$, with constraints favoring the VCG model due to its propensity to exhibit quintessence-like behavior.

On the other hand, many models of gravity have been proposed in which the Einstein-Hilbert action is modified by adding some scalars of other fields, for example, $f(R)$ gravity \cite{sotiriou2010f},
where $f$ is an arbitrary function of the Ricci scalar $R$ . Other modified theories of gravity include brane-world gravity \cite{maartens2010brane} and Tensor-Vector-Scalar (TeVeS) gravity \cite{PhysRevD.70.083509}. Moreover, if the requirement for a metric-compatible connection is relaxed, further novel modifications can be formulated.



In this context, torsion and non-metricity, besides curvature, can also represent the affine properties of a manifold. This gave rise to a theory called the "Teleparallel Equivalent to GR" (TEGR) \cite{buchdahl1970non, Aldrovandi2012TeleparallelGA}, in which the underlying gravity is described by torsion $T$. 
In \cite{jimenez2018coincident}, a modification of TEGR namely "Symmetric Teleparallel Equivalent to GR" (STEGR) was proposed in which the underlying gravitational interaction is described by the non-metricity $Q$ with no torsion and curvature. In a non-Riemannian geometry, $Q$ measures the change of vector length when it is being parallel transported. $f(Q)$ gravity \cite{jimenez2020cosmology} is an extension of the STEGR theory in which the action is described by $S=\int d^4x\sqrt{-g}f(Q)$, where $f(Q)$ is an arbitrary function of $Q$. It is important to note however, that it has been shown by \cite{paliathanasis2025telephone} that linear forms of $f(Q)$ will be essentially equivalent to STEGR/GR. Despite this, \cite{das2024cosmological} have shown that using a power law function definition for $f(Q)$, feasible solutions exist, hinting that $f(Q)$ gravity can be a feasible substitute for $\Lambda$CDM.

The authors in \cite{lazkoz2019observational} have constrained various functional forms of $f(Q)$ using cosmological observations from Type Ia Supernovae and BAO. For more recent detailed studies on $f(Q)$ gravity, we refer the reader to \cite{barros2020testing,dimakis2021quantum,khyllep2021cosmological,khyllep2022cosmology,mandal2020cosmography,mandal2020energy}. However, all of of these models assumed minimal coupling of $f(Q)$ gravity with matter. Interestingly, non-minimal coupling to matter was recently studied in \cite{harko2018coupling}, where the underlying action is $S=\int d^4x\sqrt{-g}\{f_1(Q)/2+f_2(Q)\mathcal{L}_m\}$. Here, $f_1(Q)$ and $f_2(Q)$ are two arbitrary functions of $Q$ and $\mathcal{L}_m$ is the matter Lagrangian. Assuming a perfect fluid matter distribution, observational constraints on the EOS parameters in power-law non-minimally coupled $f(Q)$ cosmology were obtained in \cite{mandal2021constraint} and the authors showed that $f(Q)$ gravity displays quintessence behavior.

Motivated by the success of the CG models, we investigate non-minimally coupled $f(Q)$ gravity with baryonic matter, radiation and a family of CG models (GCG, MCG, VCG) as the background fluid. Combining 
non-minimally coupled $f(Q)$ gravity with CG models is motivated by their complementary strengths: GCG, MCG and VCG capture a broad spectrum of cosmological behavior, while 
$f(Q)$ adds geometric effects via non-metricity. Together, they offer a framework where the universe's acceleration could arise from both matter and geometry. This synergy also enables us to study the effect of modifications to gravity on CG parameters, thus uncovering new phenomenology, which is possibly not captured by either of the two factors alone. This approach to combining the CG EOS with a modified gravity model has been previously carried out in studies such as \cite{Barreiro2004}, or \cite{Chattopadhyay2017,SultanaGudekliChattopadhyay+2024+51+70} who have used $f(T)$ models (Here, $T$ is the torsion scalar). Furthermore, it may also provide a new alternative to understanding the recent observational tensions. Finally, this combined framework enriches us with a flexible parameter space and can reveal new insights into the relationship between exotic matter and geometry.

We constrain the model parameters using three different data sets: a variety of direct measurements of the Hubble parameter $H(z)$ at different redshifts $z$ \cite{sharov2018predictions} (OHD), data from BAO measurements \cite{2012bao}, and measurements of the various properties of quasars (short for quasi-stellar objects, which we denote in the equations and figures with QSO) \cite{Risaliti_2015, Lusso_2020, lusso2020quasars_new}. The QSO data extend up to $z\sim5$, making them one of the few observational tools that effectively probes the high redshift range
$z\sim2.5-5$. We are using the QSO data because it is expected to place tighter constraints on the cosmological parameters, offering valuable insights into the dynamics of dark energy and the geometry of the universe. In \cite{khadka2020using}, the authors used OHD+BAO+QSO to set constraints on flat and non-flat versions of $\Lambda$CDM, $X$CDM and $\phi$CDM. We extend their analysis and apply these datasets on our proposed models. Notably, this work represents the first application of 
$f(Q)$ gravity tested with quasar observations.

This work is structured as follows: Sec. \ref{sec:model} introduces $f(Q)$ gravity, coupled non-minimally to the matter fields. In Sec. \ref{sec:flatflrw}, we obtain the field equations against the backdrop of a spatially flat Friedmann-Lemaitre-Robertson-Walker (FLRW) metric for the three proposed models. Sec. \ref{sec:data} details the datasets and likelihood functions used to constrain the model parameters. In Subsection \ref{sec:circularity}, we
describe the calibration of quasar measurements for model-independent cosmological analysis (following the work by \cite{Wei_2020}) and compare results across three QSO datasets: 2015 \cite{Risaliti_2015}, 2019 \cite{Lusso_2020}, and 2020 \cite{lusso2020quasars_new}. We also derive the analytic expression for errors in the comoving distance assuming a B\'ezier-style equation for the Hubble parameter.
Parameter constraints, kinematic diagnostics and information criteria are discussed in Secs. \ref{sec:results} and \ref{sec:diagnostics} to evaluate deviations from 
$\Lambda$CDM. Finally in Sec. \ref{sec:conclusions}, we summarize our main findings and their implications and discuss potential future avenues.
\section{\texorpdfstring{$f(Q)$}{f(Q)} Model}
\label{sec:model}Similar to the non-minimally coupled $f(R)$ gravity \cite{thakur2011non}, the action for non-minimally coupled $f(Q)$ gravity \cite{jimenez2018coincident,mandal2021constraint} is given by 
\begin{equation}
\label{eq:Action}
    S=\int d^4x \sqrt{-g}\left[\frac{1}{2\kappa^2}f_1(Q)+f_2(Q)\mathcal{L}_m \right]
\end{equation}
where $f_1(Q)$ and $f_2(Q)$ are functions of non-metricity $Q$, $\mathcal{L}_m$ is the matter Lagrangian density, $g\equiv \text{det}(g_{\mu\nu})$, $g_{\mu\nu}$ is the underlying metric and $\kappa^2=8\pi \mathcal{G}$ where $\mathcal{G}$ is the gravitational constant. Here, we set $\kappa^2=1$. If $f_2(Q)=1$ and $f_1(Q)=Q$ , we retrieve the standard well-studied $f(Q)$ gravity \cite{jimenez2018coincident}. 
The non-metricity Q is given as
\begin{equation}
\label{eqn:non-metricity}
    Q=g^{\mu\nu}(L\indices{^\lambda_{\sigma\lambda}}L\indices{^\sigma_{\mu\nu}}-L\indices{^\lambda_{\sigma\mu}}L\indices{^\sigma_{\nu\lambda}})
\end{equation}
Here, $L\indices{^\lambda_{\sigma\mu}}$ is the disformation tensor which is defined as
\begin{equation}
\label{eqn:disformation tensor}
    L\indices{^\lambda_{\sigma\mu}}=-\frac{1}{2}g^{\lambda\gamma}(\nabla_\mu g_{\sigma\gamma}+\nabla_\sigma g_{\gamma\mu}-\nabla_\gamma g_{\sigma\mu})
\end{equation}
The non-metricity $Q$ can be calculated as $Q=-Q_{\lambda\mu\nu}P^{\lambda\mu\nu}$, where $Q_{\lambda\mu\nu}=\nabla_\lambda g_{\mu\nu}$ and $P\indices{^\lambda_{\mu\nu}}$ is the super-potential given by
\begin{equation}
    P\indices{^\lambda_{\mu\nu}}=\frac{1}{4}g_{\mu\nu}\left(Q^{\lambda}-\Tilde{Q}^{\lambda}\right)-\frac{1}{4}\delta^{\lambda}_{(\mu}Q_{\nu)}-\frac{1}{2}L\indices{^\lambda_{\mu\nu}}
\end{equation}
where $Q_\lambda=Q\indices{_\lambda^\mu_\mu}$ and $\Tilde{Q}_{\lambda}=Q\indices{^\mu_{\lambda\mu}}$ are the two independent traces of the non-metricity tensor. The energy-momentum tensor $T_{\mu\nu}$ is specified as
\begin{equation}
    T_{\mu\nu}=-\frac{2}{\sqrt{-g}}\frac{\delta(\sqrt{-g}\mathcal{L}_m)}{\delta g^{\mu\nu}}
\end{equation}
Varying Eqn. (\ref{eq:Action}) with respect to $g_{\mu\nu}$ establishes the following modified Einstein's equation:

\begin{multline}
\label{eq:field_eqn}
    \frac{2}{\sqrt{-g}}\nabla_\lambda\left(\sqrt{-g}f_Q P\indices{^\lambda_{\mu\nu}}\right)+\frac{1}{2}g_{\mu\nu}f_1(Q)+f_Q\left(P_{\mu\lambda\sigma}Q\indices{_\nu^{\lambda\sigma}}-2Q_{\lambda\sigma\mu}P\indices{^{\lambda\sigma}_{\mu}}\right)
    =-f_2(Q)T_{\mu\nu}
\end{multline}
where $f_Q=f^{'}_1(Q)+2f^{'}_2(Q)\mathcal{L}_m$ and $\{\hspace{2pt}^{'}\hspace{2pt}\}$ represent differentiation with respect to $Q$. On the other hand, varying with respect to the connection \cite{harko2018coupling} gives
\begin{equation}
\label{connection}
    \nabla\mu\nabla\nu(\sqrt{-g}f_QP\indices{^{\mu\nu}_{\lambda}}-f_2(Q)H\indices{_{\lambda}^{\mu\nu}})=0
\end{equation}
where $H\indices{_{\lambda}^{\mu\nu}}=-\frac{1}{2}\frac{\delta(\sqrt{-g}\mathcal{L}_m)}{\delta \Gamma\indices{^{\lambda}_{\mu\nu}}}$ is the hyper-momentum tensor density and $\Gamma\indices{^{\lambda}_{\mu\nu}}$ are the Christoffel symbols. This variation can be achieved in two ways: by using inertial variation \cite{golovnev2017covariance}, where the connection is treated as a pure gauge; or by introducing Lagrange multipliers \cite{jimenez2018teleparallel} in the action to enforce vanishing curvature and torsion, while allowing for a general connection.

In the next section, we find the field equations in the spatially flat FLRW universe.  
\section{Flat FLRW universe}
\label{sec:flatflrw}
Consider the flat FLRW metric
\begin{equation}
\label{eq:metric}
    ds^2=-dt^2+a^2(t)(dx^2+dy^2+dz^2)
\end{equation}
where $a(t)$ is the scale factor. We have set the lapse function to unity by a general time reparametrization. According to \cite{harko2018coupling}, Eq. \ref{connection} is identically satisfied for
the model \ref{eq:Action} in the background Eq. \ref{eq:metric}. In the backdrop of this metric, the non-metricity, using Eqs. (\ref{eqn:non-metricity}),(\ref{eqn:disformation tensor}) is given as 
\begin{equation}
    Q=6H^2
    \label{eq:qhrel}
\end{equation}
where $H=\Dot{a}/a$ and $\{\hspace{2pt}{\Dot{}}\hspace{2pt}\}$ is the derivative with respect to cosmological time. We assume that the energy-momentum tensor $T_{\mu\nu}$ is given in the form of a perfect fluid 
\begin{equation}
\label{eq:T_mu_nu}
    T_{\mu\nu}=(p+\rho)u_{\mu}u_{\nu}+pg_{\mu\nu}
\end{equation}
where $\rho$ and $p$ are the energy density and pressure respectively. The four-velocity $u_{\mu}$ satisfies the normalization $u_{\mu}u^{\mu}=-1$. Substituting the metric Eq. (\ref{eq:metric}) and Eq. (\ref{eq:T_mu_nu}) in Eq. (\ref{eq:field_eqn}), we get the following Friedmann equations:  
\begin{equation}
\label{eq:friedman_eqns}
    \begin{split}
        3H^2&=\rho_{\text{eff}}=\frac{f_2}{2f_Q}\left(-\rho+\frac{f_1}{2f_2}\right)\\
        \Dot{H}+3H^2&=-p_{\text{eff}}=\frac{f_2}{2f_Q}\left(p+\frac{f_1}{2f_2}\right)-\frac{\Dot{f}_Q}{f_Q}H
    \end{split}
\end{equation}
Here $\rho_{\text{eff}}$ and $p_{\text{eff}}$ represent the effective energy density and pressure respectively. Using Eqs. (\ref{eq:friedman_eqns}), the continuity equation is given as  
\begin{equation}
    \Dot{\rho}+3H(p+\rho)=-6\frac{f^{'}_2}{f_2}H\Dot{H}(\mathcal{L}_m+\rho)
\end{equation}
For a perfect fluid with the Lagrangian prescription $\mathcal{L}_m=-\rho$ \cite{harko2018coupling}, we retrieve the standard continuity equation:
\begin{equation}
\label{eq:continuity}
    \Dot{\rho}+3H(p+\rho)=0
\end{equation}
This aligns with the fact that, in an isotropic and homogeneous background, the connection equation \ref{connection} becomes trivial \cite{harko2018coupling}. In this work, we assume the following simple functional forms for $f_1(Q)$ and $f_2(Q)$:
\begin{equation}
    f_1(Q)=\alpha Q^m, \hspace{10pt} f_2(Q)=Q 
\end{equation}
where $\alpha$ and $m\ne 1$ are constants. The primary motivation for adopting these forms for $f_1$ and $f_2$ is the fact that the Eqs. (\ref{eq:friedman_eqns}) are ordinary differential equations, for which power-law and exponential solutions are tractable, both of which were extensively studied in \cite{ jimenez2018coincident}. The authors found that the universe can undergo an accelerated expansion of the power-law type, contingent on the specific choice of $m$. Additionally, \cite{anagnostopoulos2023new} demonstrated that the power law model can account for the late-time acceleration of the universe without violating the BBN constraints. Another reason for this choice is that it has been shown that linear forms of $f(Q)$ will reproduce results equivalent to STEGR/GR \cite{paliathanasis2025telephone}, implying that a choice of $f_1(Q)=Q$ and $f_2(Q)=1$ for example, are unacceptable. A power-law form has been studied before in studies such as \cite{mandal2021constraint,BHARDWAJ2025170128,das2024cosmological}. Specifically, in \cite{das2024cosmological}, it has been shown that when using $f_1(Q)=\alpha Q^m$ and $f_2(Q)=Q$, certain values of $\alpha$ and $m$ can explain late-time cosmological acceleration very well. As a result, we study this form and derive constraints on $m$ for an expanding universe in the next subsection.
\subsection{Analytical Investigation for Single-Component Universe}
From the functional forms that we have introduced, we have $f_Q=\alpha mQ^{m-1}-2\rho$. Substituting this in the first Friedmann Equation (Eqn \ref{eq:friedman_eqns}), we find
\begin{equation}
    \rho = \alpha(m-0.5)Q^{m-1}.
    \label{eq:rhosimplif}
\end{equation}
From this, we deduce that the term $\alpha(m-0.5)$ must be positive.

In order to find an understanding of what our model predicts, we investigate here what our equations imply for a single-component universe whose EOS is $\rho=\rho_0a^{-3(1+\omega)}$ where $\omega$ is the EOS parameter and $\rho_0$ is the current value of density. Substituting this in Eqn. \ref{eq:rhosimplif} and rearranging, we find $Q$ as a function of the scale factor $a$:
\begin{equation}
    Q =Q_0 a^{\frac{-3(1+\omega)}{m-1}}
    \label{eq:Qomega}
\end{equation}
where $Q_0$ is
\begin{equation}
    Q_0 = \left(\frac{\rho_0}{\alpha(m-0.5)}\right)^{\frac{1}{m-1}}
    \label{eq:Q0def}
\end{equation}
After inserting Eqn \ref{eq:Qomega} into Eqn \ref{eq:qhrel}, we get
\begin{equation}
    \Dot{a}=\mathcal{C}a^l
\end{equation}
where $\mathcal{C}=\sqrt{Q_0/6}$ and $l=1-(3(1+\omega)/(2m-2))$. Integrating this simple differential equation gives us $a\propto t^{\frac{1}{1-l}}$. In order to have an expanding universe, we must have $1-l$ be positive, which since from the definition of $l$, we get
\begin{equation}
\label{eqn:m_greater_than_1}
    1-l=\frac{3(1+\omega)}{2(m-1)},
\end{equation}
This means that $m>1$ is expected for an expanding universe.

In the following, we introduce each of the three types of CG models we will be investigating and derive their respective Hubble equations. 
\subsection{Models}
\label{sec:modelquad}
For the purposes of this work, we consider the universe to be comprised of baryons $(b)$, radiation $(r)$  and one of the 3 types of CG. In this study, we did not consider the interactions of CG with baryons or radiation. We set the value of $a$ at the current epoch $t_0$ as 1. For pressureless baryonic matter, integrating Eq. (\ref{eq:continuity}) and using $a(z)=1/(1+z)$, the EOS is $ \rho_{b}=\rho_{b,0}(1+z)^3$,
where $\rho_{b,0}$ is the current value of baryonic energy density. Similarly, for radiation, the EOS is $ \rho_{r}=\rho_{r,0}(1+z)^4$. We define $\Omega_{i,0}=\rho_{i,0}/\rho_{cr,0}, (i=b,r,gcg, mcg, vcg)$ as the dimensionless density parameters that are defined by the current critical density, $\rho_{cr,0}=3H_{0}^2$.  We will set the variable $\Omega_{r,0}=0.0005$, which has been determined in studies such as \cite{universe7100362}. The motivation for using CG models is that they let us test how well they match observations, while also exploring whether the universe's evolution can be explained without assuming explicitly a dark energy component. Since both non-minimally coupled $f(Q)$ gravity and CG models independently address cosmic acceleration, we propose their combination as it can offer a promising avenue to explore how geometric modifications and exotic matter together influence the expansion of the universe, thus allowing for a richer phenomenology.

\subsubsection{Generalized Chaplygin Gas (GCG)}
\label{sec:gcg}

The EOS for Generalized Chaplygin Gas \cite{bento2002generalized} is (we denote the equations corresponding with this type of gas with $gcg$) 
$p_{gcg}=-C/\rho_{gcg}^{n_{gcg}}$, 
where $0< n_{gcg}\le 1$ and $C$ are positive parameters. Let $\rho_{gcg}(z=0)\equiv\rho_{gcg,0}$. Using the continuity equation Eq. (\ref{eq:continuity}), one can express the GCG's pressure as a function of z:
\begin{equation}
\label{eq:chaplygin_eos}
    \rho_{gcg}(z)=\rho_{gcg,0}\left(A_{gcg}+(1-A_{gcg})(1+z)^{3(1+n_{gcg})}\right)^{\frac{1}{1+n_{gcg}}}
\end{equation}
where $A_{gcg}=C/\rho_{gcg,0}^{n_{gcg}+1}$ is the scaled parameter and $A_{gcg}>0$. 

Substituting Eqn.(\ref{eq:chaplygin_eos}) in Eqn. (\ref{eq:friedman_eqns}), the expression for the Hubble parameter becomes:
\begin{equation}
\label{eq:hubble_chaplygin}
\begin{split}
    H_{gcg}(z)&=\left(\frac{2\rho}{\alpha(2m-1)6^{m-1}}\right)^{\frac{1}{2m-2}}\\
    &=\zeta(\alpha,m)\left(\Omega_{b,0}(1+z)^3+\Omega_{r,0}(1+z)^4+\Omega_{gcg,0}\left(A_{gcg}+(1-A_{gcg})(1+z)^{3(1+n_{gcg})}\right)^{\frac{1}{1+n_{gcg}}}\right)^{\frac{1}{2m-2}}\\
\end{split}
\end{equation}
where the prefactor $\zeta(\alpha,m)$ is 
\begin{eqnarray}
    \label{eq:alphadef}\zeta(\alpha,m)=H_0^{\frac{1}{m-1}}\left(\frac{6^{2-m}}{\alpha(2m-1)}\right)^{\frac{1}{2m-2}}
\end{eqnarray}
Setting $H_{gcg}(z=0)=H_0$, we get the following constraint equation 
\begin{equation}
\label{eq:normalization}    \Omega_{gcg,0}=\alpha H_0^{2m-4}6^{m-2}(2m-1)-\Omega_{b,0}-\Omega_{r,0}
\end{equation}
 This model has 5 free parameters $\Vec{P}_{gcg}=\{\alpha,m,n_{gcg},A_{gcg},\Omega_{b,0}\}$ that we could find with the help of observational data.

\subsubsection{Modified Chaplygin gas (MCG)}
\label{sec:mgc} 
The EOS for the Modified Chaplygin Gas (denoted by $mgc$) as introduced in \cite{benaoum2022accelerated} is $p_{mcg}=B\rho_{mcg}-C/\rho^{n_{mcg}}_{mcg}$, where $B$ is also a positive constant. For $B=0$, we retrieve the GCG's EOS and for $C=0$, we get a universe with a perfect fluid. On one hand, MCG's EOS offers constant negative pressure at low density to drive late-time acceleration and on the other, it can signal a radiation-dominated era with $B=1/3$ at high density. Thus, MCG's model surpasses GCG in versatility, effectively capturing the evolution of the universe to a large extent \cite{debnath2004role}.  The addition of the extra term $B\rho_{mcg}$ modifies the GCG's EOS slightly in the exponent as:
\begin{equation}
    \rho_{mcg}(z)=\rho_{mcg,0}\left(A_{mcg}+(1-A_{mcg})(1+z)^{3(1+B)(1+n_{mcg})}\right)^{\frac{1}{1+n_{mcg}}}
\end{equation}
where the rescaled $A_{mcg}=C/(\rho_{mcg,0}^{n_{mcg}+1}(1+B))$ and $\rho_{mcg,0}\equiv\rho_{mcg}(z=0)$.  Then the Hubble equation becomes
\begin{equation}
\label{eq:hubble_chaplygin_modified} 
H_{mcg}(z)=\zeta\left(\Omega_{b,0}(1+z)^3+\Omega_{r,0}(1+z)^4
+\Omega_{mcg,0}\left(A_{mcg}+(1-A_{mcg})(1+z)^{3(1+B)(1+n_{mcg})}\right)^{\frac{1}{1+n_{mcg}}}\right)^{\frac{1}{2m-2}}
\end{equation}
where $\zeta$ is the same term as in Eq. \ref{eq:alphadef}. The constraint equation is same as before:
\begin{equation}
\label{eq:normalization2}    \Omega_{mcg,0}=\alpha H_0^{2m-4}6^{m-2}(2m-1)-\Omega_{b,0}-\Omega_{r,0}
\end{equation}
Thus, here we have the 6 free parameters $\Vec{P}_{mcg}=\{\alpha,m,n_{mcg},A_{mcg},\Omega_{b,0}, B\}$.
\subsubsection{Variable Chaplygin gas}
The EOS for the Variable Chaplygin Gas (denoted by $vgc$) is $p_{vcg}=-C(a)/\rho_{vcg}$, where $C(a)$ is now a function of time.
In the works \cite{bento2002generalized, sen2002tachyon}, it was shown that VCG arises naturally from the dynamics of a generalized $d$-brane in a $(d+1,1)$ dimensional spacetime. The authors in \cite{chraya2023variable} took the following dynamics for $C$: $C(a)=C_0a^{-n_{vcg}}$, where $C_0$ and $n_{vcg}$ are constants and they showed that for an accelerated universe, $n_{vcg}<4$ and $C_0>0$. Substituting in Eqn. (\ref{eq:continuity}), the form of the density of the VCG gas becomes
\begin{equation}
\label{eqn:vcg_rho}
    \rho_{vcg}(z)=\sqrt{\frac{6}{6-n_{vcg}}C_0(1+z)^{n_{vcg}}+D(1+z)^6}
\end{equation}
where $D$ is the constant of integration. The case $n_{vcg}=0$ corresponds to the original CG model. At earlier times, the second term dominates i.e. $\rho_{vcg}\propto (1+z)^3$, giving rise to dust-like matter. However, at later times, the first term dominates leading to the scale factor $a\propto t^{4/n_{vcg}}$. The universe clearly accelerates for $n_{vcg}<4$. Rescaling by a positive parameter $A_{vcg}=D/((6C_0)/(6-n_{vcg})+D)$, Eqn. \ref{eqn:vcg_rho} becomes 
\begin{equation}   \rho_{vcg}=\rho_{vcg,0}\bigl(A_{vcg}(1+z)^6+(1-A_{vcg})(1+z)^{n_{vcg}}\bigr)^{\frac{1}{2}}
\end{equation}
where $\rho_{vcg,0}$ is the current value of $\rho_{vcg}$. Inserting this in Eqn. \ref{eq:friedman_eqns}, the Hubble equation for this model is given by
\begin{equation}
\label{eq:hubble_chaplygin_variable} H_{vcg}(z)=\zeta\left(\Omega_{b,0}(1+z)^3+\Omega_{r,0}(1+z)^4+\Omega_{vcg,0}\left(A_{vcg}(1+z)^6+(1-A_{vcg})(1+z)^{n_{vcg}}\right)^{\frac{1}{2}}\right)^{\frac{1}{2m-2}}
\end{equation}
with $\zeta$ identical to that in Eq. \ref{eq:alphadef} and the same constraint equation as follows:
\begin{equation}
\label{eq:normalization3}    \Omega_{vcg,0}=\alpha H_0^{2m-4}6^{m-2}(2m-1)-\Omega_{b,0}-\Omega_{r,0}
\end{equation}
There are 5 free parameters for this model which are $\Vec{P}_{vcg}=\{\alpha,m,n_{vcg},A_{vcg},\Omega_{b,0}\}$.
\\
\\
For comparison, our baseline model is $\Lambda$CDM with $\Omega_{m,0}=0.315$ and $\Omega_{\Lambda,0}=1-\Omega_{m,0}$ being the current values of matter density (baryons+CDM), and dark energy respectively \cite{aghanim2020planck}.
\section{Datasets and Methods}
\label{sec:data}
We will be using three different datasets in order to constrain the aforementioned parameters. These datasets are:
\subsection{Hubble Measurements}
\label{sec:hdata}
Measurements have been made of the Hubble parameter $H(z)$, its associated error $\sigma_H$, and the corresponding redshift $z$ in studies \cite{hz1,hz2,hz3,hz4,hz5,hz6,hz7,hz8,hz9,hz10,hz11,hz12,hz13,hz14,hz15,hz16,hz17,hz18,hz19}. In \cite{Solanki_2021}, the authors have compiled a dataset consisting of 57 measurements. We use the following $\chi^2$ function to constrain the free parameters
\begin{equation}
\label{eq:xi2hubble} \chi^2_{OHD}=\sum\limits_{i=1}^{57}\left\{\frac{(H_{model}(z_i,\Vec{P}_{model})-H_{data}(z_i))^2}{\sigma_{H(z_i)}^2}+\log{2\pi \sigma_{H(z_i)}^2}\right\}
\end{equation}

where $H_{model}(z_i,\Vec{P}_{model})$ and $H_{data}(z_i)$ are the theoretical and observed values of the Hubble parameter at $z=z_i$ respectively (the theoretical equations are Equations \ref{eq:hubble_chaplygin}, \ref{eq:hubble_chaplygin_modified}, and \ref{eq:hubble_chaplygin_variable}, and $\Vec{P}_{model}$ represents the free parameters of each CG model). Here, $\sigma_{H(z_i)}$ is the corresponding error in Hubble parameter at redshift $z_i$. Based on \cite{Amati_2019}, we set $H_0=67.76$ $\mathrm{km} \mathrm{s}^{-1} \mathrm{Mpc}^{-1}$. We use each of the three Hubble model equations as $H_{model}(z_i)$ to obtain a fit with the aforementioned data.
\subsection{Baryon Acoustic Oscillations}
\label{sec:bao}
Baryon Acoustic Oscillations (BAO) emerge in the early stages of the universe's evolution.
The sound horizon $r_s$ which defines the characteristic scale of BAO, is visible at the photon decoupling redshift $z^{*}$, whose expression is
\begin{equation}
    r_s(z^{*})=\frac{c}{\sqrt{3}}\int\limits_{0}^{\frac{1}{1+z^{*}}}\frac{1}{a^2H(a)\sqrt{1+a(3\Omega_{b,0}/4\Omega_{\gamma,0})}}da
\end{equation}
where $\Omega_{\gamma,0}$ is the current value of the photon density and $c$ is the speed of light. 
We take the value of $z^{*}$ to be $1091$. Before constructing the BAO observable, we first define a few key quantities, starting with the comoving angular diameter distance $D_C(z)$:
\begin{equation}
\label{eq:comoving_distance} D_C(z)=c\int\limits_{0}^{z}\frac{1}{H(z')}dz'
\end{equation}
 The dilution scale is $D_V(z)=(D_C^2(z)cz/H(z))^{1/3}$. The BAO observable is then given as $D_C(z^{*})/D_V(z_{BAO})$, whose observed values are available from the studies of \cite{2012bao, eisenstein2005detection, 2011bao}. 
Therefore, to carry out the MCMC sampling, it is necessary to define the $\chi^2_{BAO}$ function for the BAO dataset:
\begin{equation}
    \chi^2_{BAO}=X^TC^{-1}X
    \label{eq:xi2bao}  
\end{equation}
where $C^{-1}$ is the covariance matrix and $X$ is a difference column vector between the theoretically computed value of $D_C(z^{*})/D_V(z_{BAO})$ (which we find with each of our three Hubble equations, for each CG) and its corresponding observed values (given by \cite{2012bao}).
\subsection{Quasars}
\label{sec:quasardata}

Quasars are active galactic nuclei with very high persistent luminosity. They are among the furthest (and oldest) objects that can be observed in the universe. 
Their potential to investigate cosmological models gains more importance once one considers that a notable number of relatively high redshifts quasars have recently been discovered, thanks to projects such as the Sloan Digital Sky Survey (SDSS) \cite{2020sdss}. One method to use quasars in order to test the validity of various cosmological models is to use them as "standard candles"; in the same way that Cepheid Variables \cite{1929hubble}, and more recently type Ia Supernovae \cite{riess1998observational} have been used. This has become possible thanks to the work of \cite{Lusso_2020}, who have developed a technique to deduce the distance moduli of quasars. This technique is based on the observed relation between the X-ray and Ultraviolet luminosity of quasars \cite{tananbaum1979}.

In 2015, Risaliti and Lusso published a quasar dataset of $808$ quasars \cite{Risaliti_2015} spanning the redshift range $0.061\le z \le 6.28$. This data table consists of the redshifts ($z$), the X-ray fluxes, and ultraviolet fluxes of these respective quasars (denoted in the text by $\log F_{X}$ and $\log F_{UV}$ respectively). However, the measured intrinsic dispersion $\sigma_{\text{int}}$ in quasar's UV and X-ray luminosity relation was relatively high, at $0.320\pm0.008$. To address this, the authors \cite{Lusso_2020} compiled $1598$ quasars in the range $0.04\le z\le5.1$ from multiple sources. This decreased the dispersion to $0.230\pm0.004$.  These datasets are referred to in this work as the 2015 and 2019 datasets, respectively.
Finally, in 2020,  a table of $\approx 2400$ quasars upto $z\sim7.5$ \cite{lusso2020quasars_new} was assembled, where $\sigma_{\text{int}}$ does not evolve with redshift and was measured to be the lowest at $0.21\pm0.06$ and we call this the 2020 dataset. These tables have been provided in the VizieR catalog \cite{vizier}. By leveraging the well-established X-ray-UV luminosity relation \cite{Risaliti_2015}, their distance moduli can be determined. In this paper, we calibrate these datasets using a method that is independent of any specific cosmological model.
\subsubsection{Cosmology-Independent Calibration}
\label{sec:circularity}
A key challenge in utilizing these datasets for cosmological models beyond $\Lambda$CDM lies in the fact that the methodology employed by \cite{Lusso_2020} to derive distance moduli assumes a $\Lambda$CDM framework. This issue, commonly referred to as the circularity problem, can be addressed by constructing curvature-dependent luminosity distances, through the use of cosmic-chronometer measurements \cite{Wei_2020}. These measurements, obtained from passively evolving galaxies, provide $H(z)$ data points that are independent of any cosmological model \cite{2002ApJ...573...37J}. Using this approach, \cite{Wei_2020} calibrated the distance moduli of the observations with other models. We have incorporated their method (with some modifications) in this section which is explained below. 

The B\'ezier fit  proposed by \cite{Amati_2019, Wei_2020} has been used to describe the Hubble parameter $H(z)$ as a function of redshift, which is based on different $31$ cosmology-independent measurements of $H(z)$ in different redshifts. This function is used to calibrate the quasar datasets. In order to avoid an oscillating $H(z)$ fit, the authors in \cite{Amati_2019} have used the following form of the equation:\\
\begin{equation}
H(z) = \beta_0\left(1-\frac{z}{z_m}\right)^2 + 2\beta_1\left(1-\frac{z}{z_m}\right)\left(\frac{z}{z_m}\right) + \beta_2\left(\frac{z}{z_m}\right)^2 
\label{eq:hubble-nogeo}
\end{equation}

where $\beta_d$ are the coefficients of the Bernstein polynomial and $z_m$ is the maximum measured redshift in the 31 measurements which is $z_m=1.965\approx2$. 

To circumvent the circularity problem, we first start with the X-ray - UV relation \cite{Risaliti_2015}:

\begin{equation}
    \log{F_{X,theory}} =  \gamma \log{F_{UV}} + \beta ' +2(\gamma -1)\log(D_L^{theory}(z,\Omega_k))
    \label{eq:fxmlf}
\end{equation}
where $F_X$ and $F_{UV}$ are the X-ray and Ultraviolet fluxes respectively, $\beta^{'}$ and $\gamma$ are two unknown parameters that characterize the X-ray/UV relation, and $D_L^{theory}$ is the luminosity distance obtained from theory and is given by the following equations:
\begin{equation}
D_L^{theory}(\Omega_k,z)=
    \begin{cases}
    (1+z) \frac{D_H}{ \sqrt{\Omega_k}}\sinh{\left[\frac{\sqrt{\Omega_k}D_C(z)}{D_H}\right] }, & \text{for } \Omega_k>0\\[10pt]
    (1+z) D_C(z) , & \text{for } \Omega_k=0 \\[10pt]
    (1+z) \frac{D_H}{ \sqrt{\Omega_k}}\sin{\left[\frac{\sqrt{\lvert \Omega_k \lvert}D_C(z)}{D_H}\right] }, & \text{for } \Omega_k<0\\[10pt]
    \end{cases}
    \label{eq:lumdis}
\end{equation}
where $\Omega_k$ is the curvature parameter, $D_H=c/H_0$, and $D_C(z)$ is the co-moving distance as defined in Eq. \ref{eq:comoving_distance}. Since  this integral is solvable for the Hubble expression we defined here with Bernstein polynomials (Eq. \ref{eq:hubble-nogeo}), we obtain and use its analytical expression in  \ref{app:dcal} instead of numerical integration. 

Then, the maximum likelihood function (MLF) can be constructed using the following definition:\\
\begin{equation}
MLF=-0.5\left\{\sum_i\left(\frac{\log{F_{X,theory-i}}-\log{F_{X,data-i}}}{\sigma_{tot,i}}\right)^2 + \log{2\pi \sigma_{tot,i}^2}\right\}
\label{eq:xi2circqso}
\end{equation}
where
\begin{equation}
\sigma_{tot,i}^2 = \sigma_{int}^2 + \sigma_{\log_{10}{F_{X,data-i}}}^2+\left[\frac{2(\gamma-1)}{\log_e{10}}\frac{\sigma_{D_L^{theory-i}}}{D_L^{theory-i}}\right]^2
\label{eq:sigma}
\end{equation}
in which $\sigma_{int}$ is the internal dispersion to tackle Eddington bias, and is another unknown parameter. Since we aren't assuming anything about the curvature parameter either; both $\Omega_K$ and $\sigma_{int}$ are assumed as free parameters. Also, $\sigma_{D_L^{theory-i}}$ is the uncertainty in luminosity distance; whose expressions we take from \cite{Wei_2020}:
\begin{equation}
\sigma_{D_L^{theory}}(\Omega_k,z)=
    \begin{cases}
(1+z)\cosh{\left[\frac{\sqrt{\Omega_k}\sigma_{D_C}}{D_H}\right] }, & \text{for } \Omega_k>0\\[10pt]
    (1+z) \sigma_{D_C} , & \text{for } \Omega_k=0 \\[10pt]
    (1+z)\cos{\left[\frac{\sqrt{\lvert \Omega_k \lvert}\sigma_{D_C}}{D_H}\right] }, & \text{for } \Omega_k<0\\[10pt]
    \end{cases}
\end{equation}
Where $\sigma_{D_C}$ is the uncertainty in the comoving distance. Unlike \cite{Wei_2020} who estimated this error using Monte Carlo simulations, we have used error propagation to find an analytical expression for $\sigma_{D_C}$ as derived in \ref{app:dcal}.
\\In order to constrain the 4 parameters $\beta'$, $\gamma$, $\sigma_{int}$, and $\Omega_k$; we once again use Equation \ref{eq:hubble-nogeo} in order to obtain luminosity distances that we could input into Equation \ref{eq:fxmlf}. By following the outlined procedure and using the data in the three mentioned datasets, we can now determine the values of the four unknown parameters. Subsequently, we could use \ref{eq:fxmlf} with the obtained values for $\beta'$ and $\gamma$ to work with any other cosmological model; since they have been obtained without the assumption of any prior model.\\
One noteworthy matter is that since what exists in the catalogues isn't luminosity distance; rather the distance modulus $\mu$, we use the following standard relation to replace $D_L$ with $\mu$:
\begin{equation}
    \label{eq:mudl}
    \mu = 5\log{D_L^{theory}(z,\Omega_k)}+25
\end{equation}
To perform this calibration, the Markov-Chain Monte Carlo code EMCEE \cite{emcee2013} has been employed to constrain these unknown parameters with Eq. \ref{eq:xi2circqso} as the MLF. We adopted uniform priors for each of the parameters. The results for the three datasets can be seen in Figure \ref{fig:3quasars} in \ref{app:comparison_qso}. In order to constrain the free parameters of our three models with the help of the quasars dataset, we make use of Eq. \ref{eq:fxmlf}. Since the $f(Q)$ models we derived were based on the assumption that $\Omega_k=0$, we use the 2019 quasar dataset to constrain the free parameters, as the value we find for $\Omega_k$ lies within $3\sigma$ of our assumption. 

We denote the values found for $\beta'$ and $\gamma$ with this analysis as $\beta'_{calib}$ and $\gamma_{calib}$ (calib: calibrated values using B\'ezier fit). With these values we can compute the calibrated distance moduli from the observations:
\begin{equation}
    \mu_{calib}(z)= \frac{5}{2(\gamma_{calib}-1)}(\log{F_{X,data-z}} -  \gamma_{calib} \log{F_{UV,data-z}} + \beta'_{calib})+25
\end{equation}
where $\{,data-z\}$ is the corresponding data value at specific $z$. The $\chi^2$ definition for Quasars (denoted with QSO) is then
\begin{equation}
\label{eq:xi2qso} \chi^2_{QSO}=\sum\limits_{i} \left\{\frac{(\mu_{model}(z_i,\Vec{P}_{model})-\mu_{calib}(z_i))^2}{\sigma_{\mu(z_i)}^2}+ \log{2\pi \sigma_{\mu(z_i)}^2}\right\}
\end{equation}
where $\mu_{model}(z_i,\Vec{P}_{model})$ is calculated first by finding the comoving distance $D_C$ from numerically integrating \ref{eq:comoving_distance}, and then using \ref{eq:lumdis} to find the luminosity distance $D_L$ and then finally turned to distance modulus with Eq. \ref{eq:mudl}. Finally, $\sigma_{\mu(z_i)}$ is the error in distance modulus whose expression we take from \cite{Wei_2020}. 

\subsection{Combined Analysis: OHD+QSO+BAO}
In order to investigate how all three sets of data affect the models' parameters, we calculate the following $\chi^2$ definition
\begin{equation}
    \chi^2_{total}=\chi^2_{OHD}+\chi^2_{BAO}+\chi^2_{QSO}
\end{equation}
where $\chi^2_{OHD}$, $\chi^2_{BAO}$, and $\chi^2_{QSO}$ are given by Eq. (\ref{eq:xi2hubble}), Eq. (\ref{eq:xi2bao}),   and Eq. (\ref{eq:xi2qso}) respectively. 
\section{Results}
\label{sec:results}
\begin{figure}[ht!]
	\centering                            
	\includegraphics[width=12cm,height=12cm]{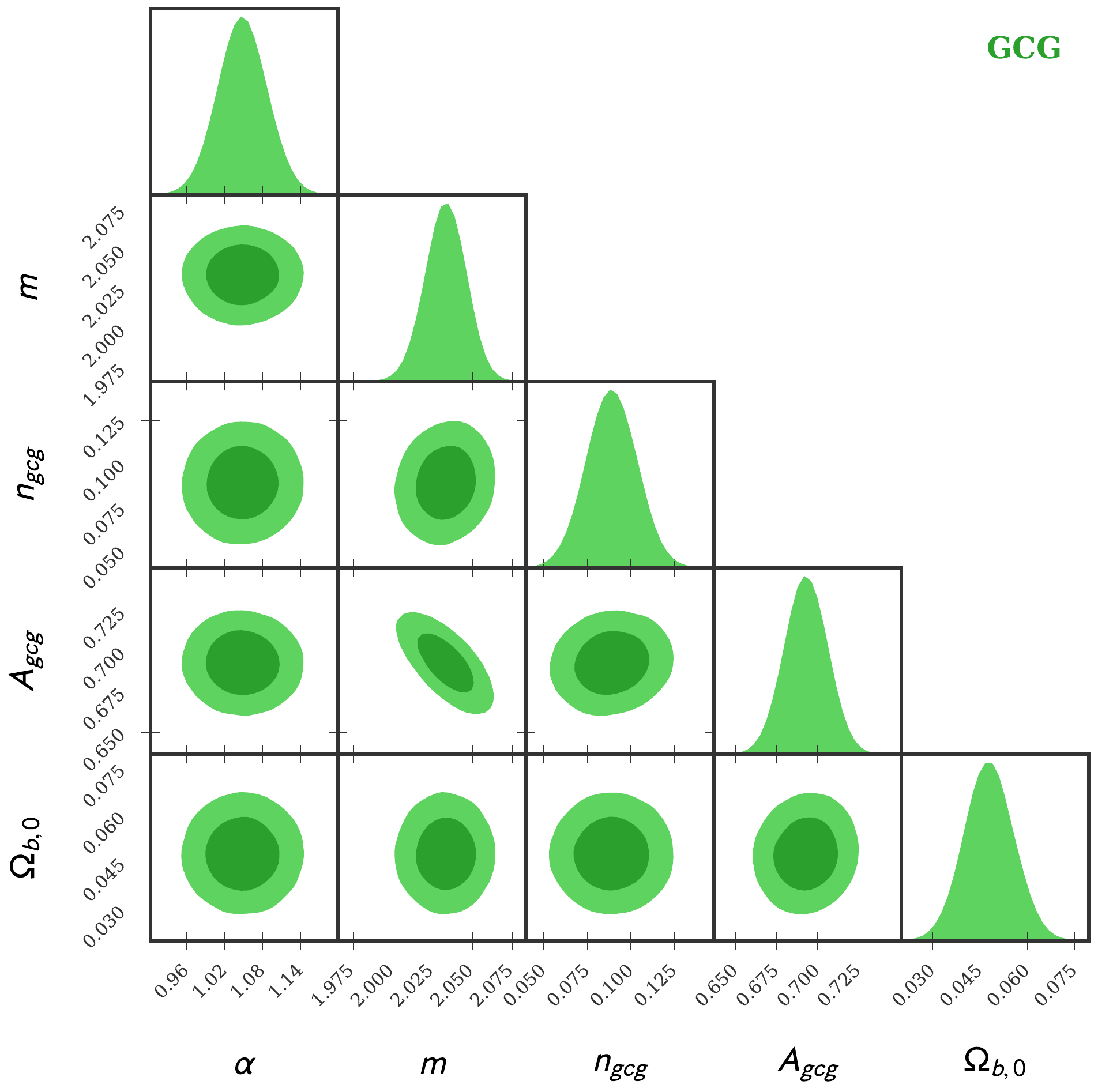}
	\caption{The corner plot above displays one-dimensional marginalized distributions and the two-dimensional contour plots for the free parameters of the GCG model with $1-\sigma$ and $2-\sigma$ error bands obtained with EMCEE using OHD+BAO+QSO. }
		\label{fig:corner_1}
\end{figure}

\begin{figure}[ht!]
	\centering                            
	\includegraphics[width=12cm,height=12cm]{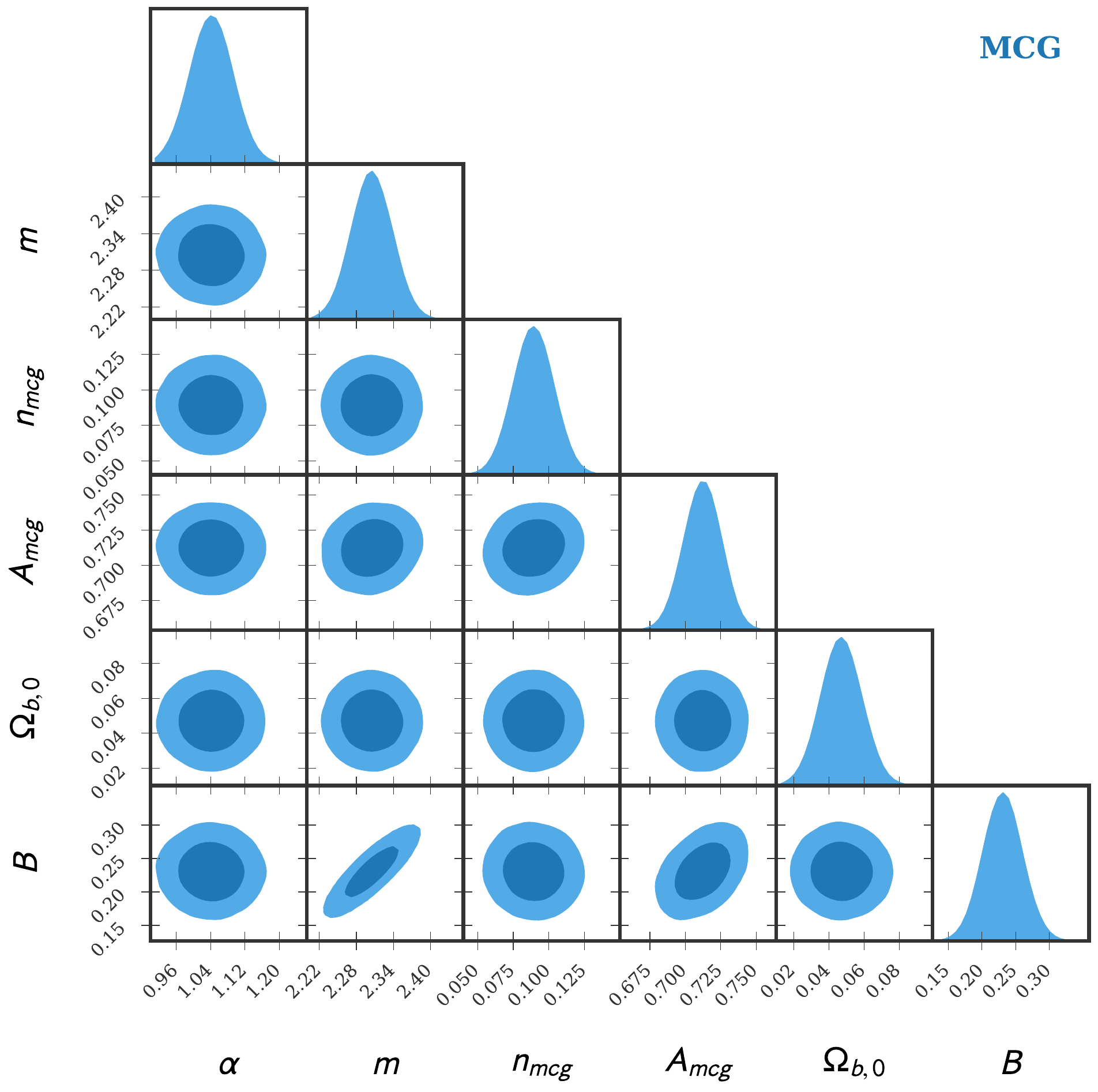}
	\caption{The corner plot above displays one-dimensional marginalized distributions and the two-dimensional contour plots for the free parameters of the MCG model with $1-\sigma$ and $2-\sigma$ error bands obtained with EMCEE using OHD+BAO+QSO.}
		\label{fig:corner_2}
\end{figure}

\begin{figure}[ht!]
	\centering                            
	\includegraphics[width=12cm,height=12cm]{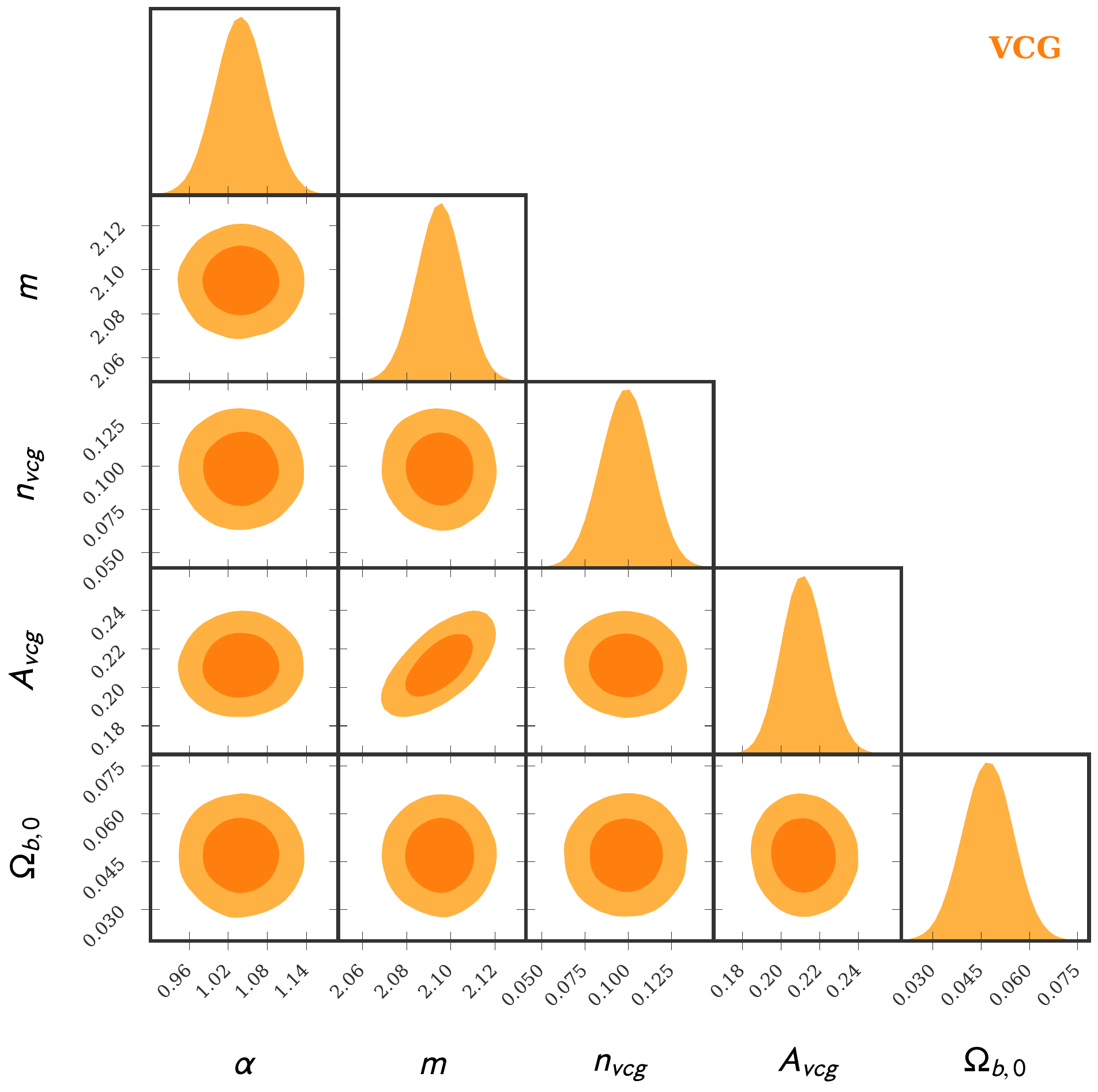}
	\caption{The corner plot above displays one-dimensional marginalized distributions and the two-dimensional contour plots for the free parameters of the VCG model with $1-\sigma$ and $2-\sigma$ error bands obtained with EMCEE using OHD+BAO+QSO.}
		\label{fig:corner_3}
\end{figure}

\begin{figure}
    \centering    \includegraphics[width=\linewidth]{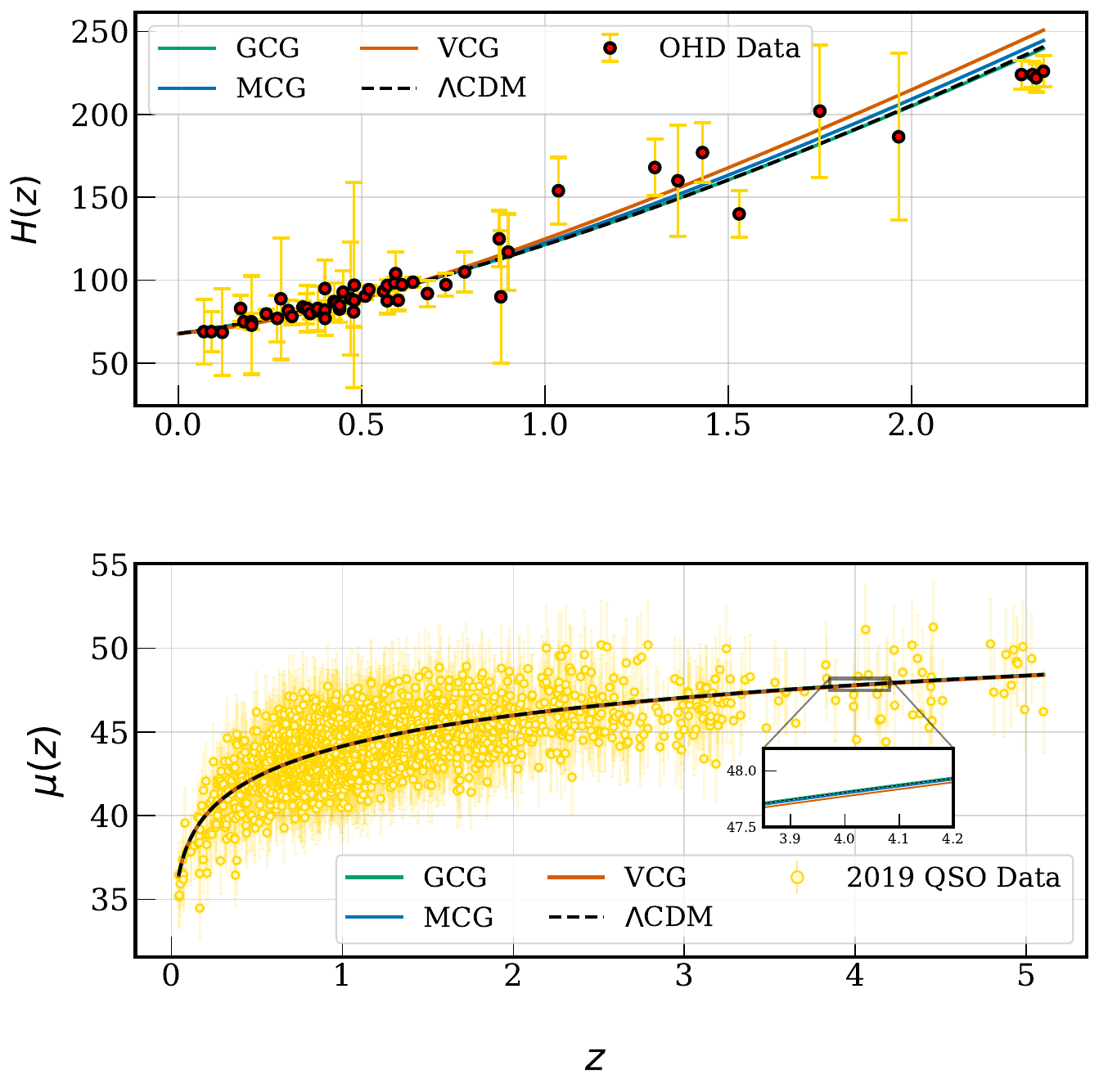}
    \caption{(a) Hubble parameter against redshift $z$. The solid line represents EMCEE fit of the proposed model to the $57$ datapoints of $H(z)$ dataset. $\Lambda$CDM model is indicated with the dashed-dot line. (b) Distance modulus $\mu(z)$ as a function of $z$. The circular scatter marker represents the EMCEE fit of the proposed model to the 2019 QSO data. $\Lambda$CDM model is indicated with the black dashed line.}
    \label{fig:fit_results}
\end{figure}

We once again use EMCEE to find the free parameters of each CG model ($P_{gcg}$, $P_{mcg}$, and $P_{vcg}$). As mentioned above, we have assumed zero spatial curvature $(\Omega_k=0)$ for our simulations in this section and have used a Gaussian distribution for the priors. With the calibrated parameters obtained in \ref{sec:circularity}, we can now use the QSO data coupled with OHD and BAO datasets to constrain our three models which are parameterized with Eqns. \ref{eq:hubble_chaplygin}, \ref{eq:hubble_chaplygin_modified}, and \ref{eq:hubble_chaplygin_variable}. The results of the simulations can be seen in the form of the corner plots in Figures \ref{fig:corner_1}, \ref{fig:corner_2}, and \ref{fig:corner_3} respectively. Figures \ref{fig:fit_results}-a and \ref{fig:fit_results}-b depict the comparison of the Hubble and Quasar datasets, for the three models against $\Lambda$CDM using the parameters obtained from the fit.

\subsection{Discussion}
\label{sec:disc}
The parameter values obtained using EMCEE are summarized in Table \ref{tab:constrained_both}. For all three models, we find $m>1$, consistent with Eq. \ref{eqn:m_greater_than_1}, required for an expanding universe. Furthermore, the $f(Q)$ parameter $\alpha$ for MCG and VCG remains within $1-\sigma$ of the value of the GCG model. Finally, we can observe that $\Omega_{b,0}$ remains consistent across the three proposed models. We now compare the constraints on the specific CG parameters with previous findings.

In the context of the GCG model, we determine that $A_{gcg}$ is positive, which indicates a positive energy density. In particular, the positivity of $A_{gcg}$ is a crucial requirement to ensure the stability of GCG perturbations, as previously established in \cite{xu2012revisiting}. In a recent paper \cite{gadbail2023cosmology}, where the authors constrained Viscous Generalized Chaplygin Gas (VGCG) in minimally coupled $f(Q)$ gravity, they derived the value for $n_{gcg}$ as $0.099\pm0.010$ at $H_0=69 $ $\mathrm{kms^{-1}Mpc^{-1}}$, by combining datasets from OHD, BAO and a Pantheon sample of $1048$ SNe Ia measurements. Their value for $n_{gcg}$ exhibits strong agreement ($1-\sigma$) with our value. In another pertinent study \cite{gadbail2022generalized}, GCG was examined as a background fluid in $f(Q,T)$ gravity, where $T$ denotes trace of energy-momentum tensor. Using OHD+SNeIa+BAO, their value of $n_{gcg}=0.0769\pm 0.0099$ for Model-I also lies within $1-\sigma$ confidence level. We conclude that there appears to be a broader trend of concordance in the estimated values of $n_{gcg}$ when constraining GCG within the $f(Q)$ framework. In Fig. \ref{fig:fit_results}-a, we observe that the $H(z)$ vs $z$ plot for GCG aligns perfectly with $\Lambda$CDM at both low and high $z$.  

On first glance, it may appear that MCG's parameters are similar to those of GCG's ones. However, the value of $B$ is more than $5-\sigma$ away from $0$. This means that the exponent $(1+B)>0$ in Eqn. \ref{eq:hubble_chaplygin_modified} drives up the effective density and hence the value of Hubble parameter, at high $z$, as is evident in both Figs. \ref{fig:diagnostics}-(a) and \ref{fig:fit_results}-(a) respectively. This indicates significant departure from $\Lambda$CDM. Now, the MCG model has never been constrained in $f(Q)$ gravity before and hence we look for constraints in regular gravity for comparison. In the paper \cite{2022_Zheng}, the authors constrained MCG in standard gravity using quasars. Our results agree with their values for $B=0.12^{+0.26}_{-0.21}$, $A_{mcg}=0.81^{+0.06}_{-0.09}$ and $n_{mcg}=0.20^{+0.58}_{-0.39}$ at $1-\sigma$ when constrained using QSO+SNeIa. 

Finally for VCG, we compare our constraints with the work \cite{chraya2023variable} where the authors used gravitational wave merger events GWTC-3 datasets to constrain VCG in Einstein gravity and obtained $A_{vcg}=0.130\pm0.079$ and $n_{vcg}=1.025\pm1.120$ which agree with our values within $1-\sigma$ confidence level in the case of Gaussian priors but disagree when compared against the combined fit of GRBs (Gamma-Ray Bursts)+GWTC-3. In \ref{fig:fit_results}-(a), we observe that VCG follows MCG and GCG at low $z$ but starts to grow apart strongly at $z>2$, again signifiying departure from the $\Lambda$CDM model. 
In Fig. \ref{fig:fit_results}-b, the QSO data at low redshifts for all three models aligns well with the flat $\Lambda$CDM model. The inset in the figure shows that VCG does start to fall below $\Lambda$CDM starting from moderate-high $z$ values.
\def\arraystretch{1.5}
\begin{table}[ht!]
    \centering
    \resizebox{\textwidth}{!}{%
    \begin{tabular}{ccccccc}
    \hline
    Model & $\alpha$ & $m$ & $n_i$ & $A_i$ & $\Omega_{b,0}$& $B$\\
    \hline 
    GCG   & $1.048^{+0.037}_{-0.037}$  &$2.033^{+0.012}_{-0.012}$ & $0.089^{+0.014}_{-0.014}$& $0.693^{+0.012}_{-0.012}$&$0.048^{+0.008}_{-0.007}$ & -\\ 
        MCG   & $1.042^{+0.049}_{-0.049}$  &$2.305^{+0.032}_{-0.032}$ & $0.089^{+0.014}_{-0.014}$& $0.712^{+0.013}_{-0.013}$&$0.047^{+0.011}_{-0.011}$&$0.231^{+0.028}_{-0.028}$ \\ 
   VCG   & $1.039^{+0.038}_{-0.037}$  &$2.095^{+0.010}_{-0.010}$ & $0.099^{+0.014}_{-0.014}$& $0.211^{+0.011}_{-0.010}$ &$0.047^{+0.007}_{-0.008}$&- \\ 
    \hline
    \end{tabular}}
    \caption{Summary of MCMC results for the free parameters of each of the 3 proposed Hubble equations using the combined datasets (QSO+OHD+BAO). The subscript `i' stands for gcg/mcg/vcg.}
\label{tab:constrained_both}
\end{table}

\subsection{Information Criteria}
\label{sec:info_crit}

In order to compare the statistical performance of the three models, we employ the Akaike Information Criterion corrected for small sample sizes (AIC) \cite{Akaike1974,liddle2007information} and Bayesian Information Criterion (BIC) \cite{1978Schwarz} given by

\begin{equation}
    \begin{split}        \text{AIC}&=\chi^2_{min}+2K+\frac{2K(K+1)}{N-K-1},\\
    \text{BIC}&=\chi^2_{min}+K\ln{N}
    \end{split}
\end{equation}

where $\chi^2_{min}$ is the minimum value of $\chi^2_{total}$, $K$ represents the number of model parameters and $N$ is the number of data points. In our combined dataset analysis, the value of $N$ is $1660$. We evaluated these criteria relative to the $\Lambda$CDM model, which we will use as a reference ($\Delta \text{AIC}=|\text{AIC}_{\text{model}}-\text{AIC}_{\Lambda \text{CDM}}|$ and $\Delta \text{BIC}=|\text{BIC}_{\text{model}}-\text{BIC}_{\Lambda \text{CDM}}|$ where model refers to GCG, MCG or VCG). For the Akaike Information Criterion (AIC), values of 
$0\le\Delta\text{AIC}<2$ indicate strong support for the proposed model over $\Lambda$CDM, $4\le\Delta\text{AIC}<7$ suggest moderate support, and $7\le\Delta\text{AIC}<10$ imply weak support. Similarly, for the Bayesian Information Criterion (BIC), $0\le\Delta\text{BIC}<2$ favors the proposed model over $\Lambda$CDM, while $2\le\Delta\text{BIC}<6$ provides evidence against it, and $6\le\Delta\text{AIC}<10$ indicates strong opposition \cite{Liddle_2004}. These values are shown in Table \ref{tab:info_crit} along with the values of  $\chi^2_{min}$ and  $\chi^2_{min}/dof$, where $dof=N-K$ represents degrees of freedom.

\begin{table}[ht!]
    \centering
    {%
    \begin{tabular}{cccccccc}
    \hline
    Model & $K$ & $\chi^2_{min}$ & $\chi^2_{min}$/dof & AIC & BIC & $\Delta$AIC &  $\Delta$BIC\\
    \hline 
    GCG & $5$ & $879.74$ & $0.53$ & $889.78$ & $916.81$ & $4.77$ & $20.98$\\ 
    MCG & $6$ & $889.53$ & $0.54$ & $901.58$ & $934.02$ & $16.57$ & $38.19$\\ 
    VCG & $5$ & $908.77$ & $0.55$ & $918.81$ & $945.84$ & $33.80$ & $50.01$\\ 
    $\Lambda$CDM & $2$ & $881.00$ & $0.53$ & $885.01$ & $895.83$ & - & -\\ 
    \hline
    \end{tabular}}
    \caption{Values of $\chi^2$, AIC and BIC for each model, as well as $\Lambda$CDM.}
\label{tab:info_crit}
\end{table}

The values of $\chi^2_{min}/dof$ for the three models and $\Lambda$CDM are relatively close. Based on the criterion for $\Delta$AIC discussed above, there is moderate support for the GCG model compared to $\Lambda$CDM, while the MCG and VCG models are disfavored. Furthermore, the fact that $\Delta \text{BIC} >10$ for all three cases is consistent with expectations, as the BIC disfavors models with additional free parameters more severely (which scales with the size of the dataset). 
\section{Diagnostics}
\label{sec:diagnostics}
\begin{figure}[ht]
    \centering
    \includegraphics[width=\linewidth]{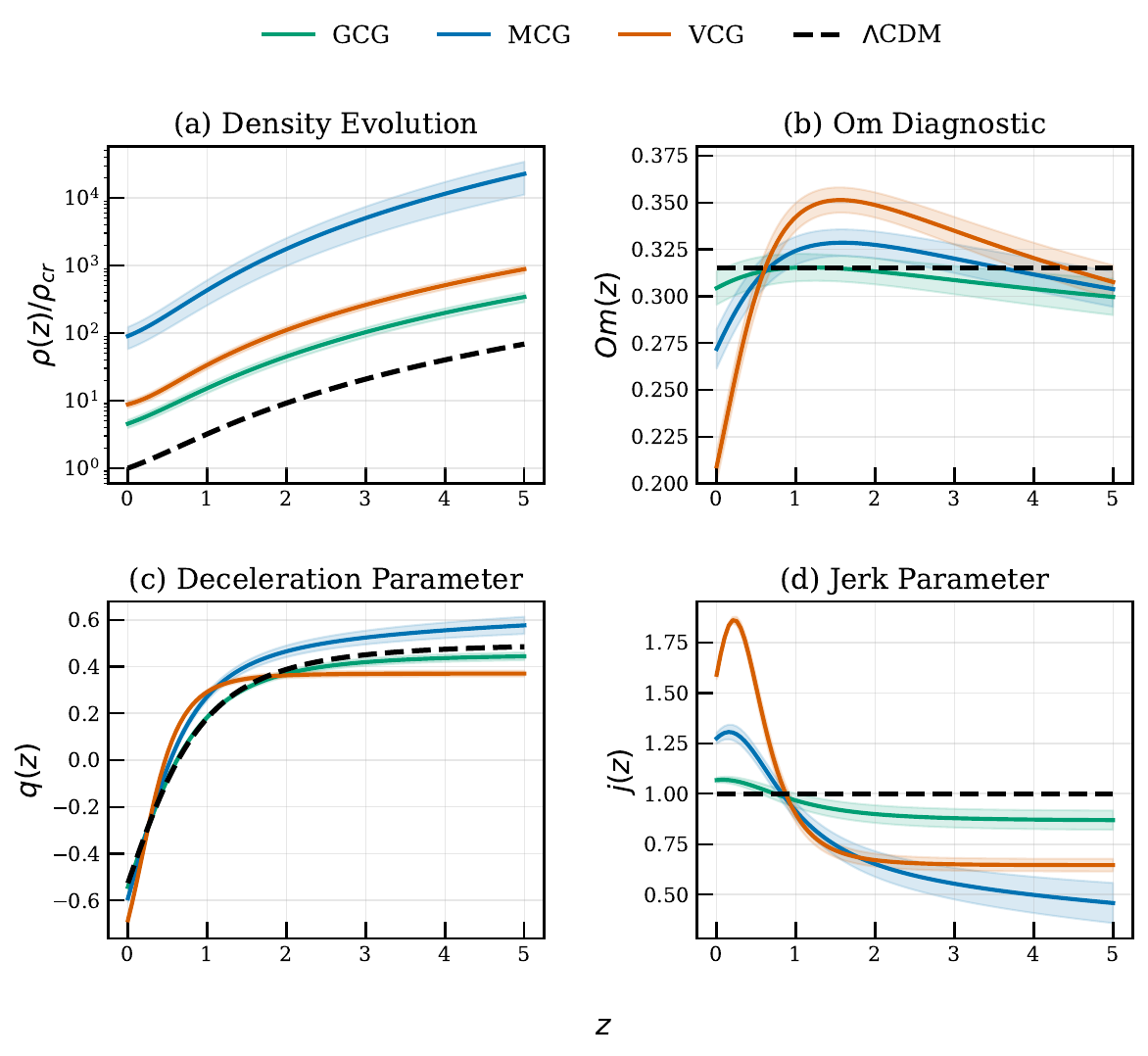}
    \caption{Plots of various diagnostics against redshift $z$ for our three models with the shaded regions representing the $1-\sigma$ uncertainty bands. Overall, the GCG model exhibits a behavior closely resembling that of the $\Lambda$CDM paradigm, with the overall strongest deviations observed in the MCG model.}
    \label{fig:diagnostics}
\end{figure}
We now examine whether our proposed model can account for the accelerated expansion of the universe by analyzing kinematic diagnostics like the deceleration parameter $q$ and the jerk parameter $j$ and additionally evolution of density and Om diagnostic. It is important to note that throughout this analysis, any parameter denoted with the subscript zero refers to its present-time value.  
\subsection{Density Parameter}
We track the evolution of density as a function of redshift. For our family of models, the Hubble equation has the form $H(z)=\zeta(\phi_1(z))^{1/(2m-2)}$. The density parameter is given by $\rho(z)/\rho_{cr}=\phi_1(z)$. As illustrated in Fig. \ref{fig:diagnostics}-(a), all three models exhibit a monotonically increasing density parameter with $z$. Notably, the MCG model demonstrates the most pronounced growth, surpassing the VCG and GCG models by at least an order of magnitude at $z=4$. Encouragingly, the effective density remains positive for all three models at $z>0$. In fact, it is strictly positive at the current time, whereas, it vanishes for $\Lambda$CDM. A comparative analysis with the findings of \cite{gadbail2023cosmology}, where a viscous-GCG (VGCG) model within minimally coupled $f(Q)$ gravity was constrained using the OHD+BAO+SNe Ia dataset at $H_0=69$ $\mathrm{kms^{-1}Mpc^{-1}}$, reveals a discrepancy: our GCG model predicts a density at 
$z=4$ that is an order of magnitude higher than their estimate of $\sim10$. After a thorough analysis, we determined that our model's $\Omega_{gcg,0}$ is $14$ times larger than that of the VGCG model presented in their work. This arises partly from a higher $m$ and the effects of non-minimal coupling, resulting in higher values for the density parameter.

\subsection{Om Diagnostic}
The Om Diagnostic, first introduced in \cite{sahni2008two}, is essentially a null test of the $\Lambda$CDM hypothesis. It helps to distinguish between dynamical dark energy models and $\Lambda$CDM with or without reference to the matter density. For a spatially flat universe, it is defined as  
\begin{equation}
    Om(z)=\frac{H^2(z)/H_0^2-1}{(1+z)^3-1}
\end{equation}
Since it is dependent only on the first derivative of the scale factor, it can be readily determined from observations. In case of $\Lambda$CDM, $Om(z)=\Omega_{m,0}=0.315$ is a constant if radiation is ignored. The cases $Om(z)>\Omega_{m,0}$ and $Om(z)<\Omega_{m,0}$ represent quintessence and phantom respectively. At higher redshifts $z\ge3.5$, we can observe from Fig. \ref{fig:diagnostics}-(b) that the $Om(z)$ diagnostic for our three models converges to the $\Lambda$CDM's prediction within $1-\sigma$. Initially exhibiting quintessence-like behavior (negative slope) up to $z\approx0.7$, the models transition beyond this point to a phantom-like regime (poitive slope), driving an increasingly accelerated cosmic expansion. Notably, a similar trend in evolution was reported in \cite{mandal2023cosmological}, where they explored modified 
$f(Q)$ gravity in conjunction with the Chevallier-Polarski-Linder (CPL) parametrization for dark energy. For our VCG model, the slope is the steepest at $z\le 1$, suggesting a significantly faster expansion rate compared to the other models. Similarly, as shown in \cite{chraya2023variable}, when the VCG model is constrained using Pantheon+GWTC-3+GRBs within the framework of Einstein's GR, it also displays phantom-like behavior in the present epoch, further reinforcing our model's predictions. 

\subsection{Deceleration Parameter}
\label{sec:q}

The deceleration parameter $q$ indicates whether the universe undergoes an accelerated expansion or not. If $q<0$, it indicates an accelerated expansion. An initial decelerating phase is required for the process of structure formation; whereas an accelerating phase in late time can explain the current observations of an accelerating expansion. 
\begin{table}[ht]
    \centering
    \begin{tabular}{cccccc}
    \hline 
    Model & $q_0$ & $j_0$ & $z_t$ \\
    \hline 
    GCG &  $-0.544_{-0.013}^{+0.013}$ & $1.068_{-0.013}^{+0.013}$ & $0.620_{-0.017}^{+0.018}$\\
    MCG & $-0.592_{-0.016}^{+0.016}$ & $1.275_{-0.028}^{+0.027}$ & $0.537_{-0.017}^{+0.017}$\\
    VCG  &$-0.687_{-0.012}^{+0.012}$ & $1.588_{-0.022}^{+0.022}$ & $0.470_{-0.012}^{+0.012}$\\
    \hline 
    \end{tabular}
    \caption{The values founded for the current deceleration $q_0$, the current jerk parameter $j_0$, and the transition redshift $z_t$ in the deceleration parameter  are listed in this table.}
    \label{tab:diagnostics_values}
\end{table} 
This indicates that a phase transition $q=0$ must occur at transition redshift $z_t$. Deceleration $q$ can be defined in terms of the Hubble Parameter as
\begin{equation}
    q(z)=-1+\frac{(1+z)}{H(z)}\frac{dH(z)}{dz}
\end{equation}
The general form of $q(z)$ for our family of models is
\begin{equation}
    q(z)=-1+\frac{3}{2(m-1)}(1+z)^3\frac{\phi_2(z)}{\phi_1(z)}
\end{equation}
where $\phi_1$ and $\phi_2=\Dot{\phi}_1/(3(1+z)^2)$ are functions of $z$ depending on the Hubble equation. Their functional forms for different models are listed in \ref{app:diagnostics}.


The values of $q_0$ and $z_t$ are tabulated in table \ref{tab:diagnostics_values}. In case of the baseline $\Lambda$CDM, $q_0=-0.53$ and $z_t=0.63$. For our suggested GCG model, the combined fit predicts a transitional redshift close to $\Lambda$CDM within $1-\sigma$, as is also evident from Fig. \ref{fig:diagnostics}-(c). In contrast, both MCG and VCG models deviate from $\Lambda$CDM in their respective $z_t$ values to more than $4-\sigma$. Nonetheless, all models consistently predict a late-time transition to accelerated expansion.

\subsection{Jerk Parameter}
\label{sec:j}

The jerk parameter $j$ is the fourth term in the Taylor series expansion of the scale factor about its present value. It is another kinematic diagnostic that measures deviations from the $\Lambda$CDM model. One can write the jerk parameter $j$ in terms of the deceleration $q$: 
\begin{equation}
    j(z)=q(z)+2q(z)^2+(1+z)\Dot{q}(z)
\end{equation}
Its value for $\Lambda$CDM universe is 1 and is independent of the redshift. For our models, $j(z)$ is
\begin{equation}
    \begin{split}
        j(z)&=1+\frac{3}{2(m-1)}(1+z)^4\frac{\phi_3(z)}{\phi_1(z)}+\frac{9(2-m)}{2(m-1)^2}(1+z)^6\frac{\phi_2^2}{\phi_1^2}
    \end{split}
\end{equation}
where $\phi_3(z)=\Dot{\phi}_2(z)$. The equations for $\phi_3$ are mentioned in \ref{app:diagnostics}. The respective model values found for $j(z=0)\equiv j_0$ are listed in table \ref{tab:diagnostics_values}. From Figure \ref{fig:diagnostics}-(d) we observe that the jerk for the GCG model is within $2-\sigma$ of the $\Lambda$CDM model and is almost flat across redshift. Around $z\approx 1$, values of the jerk for MCG and VCG are in strong agreement with $\Lambda$CDM's value, signaling a gradual acceleration of the universe's expansion. At their respective peaks, which occur near the current epoch, the expansion is accelerating at its fastest rate. After this point, the expansion reaches a more steady pace. In the CPL paper referenced in the Om diagnostic \cite{mandal2023cosmological}, their model's jerk plot exhibits a similar shape as our models' around $z=0$. 
\section{Conclusion}
\label{sec:conclusions}
In this work, we investigated the $f(Q)$ cosmology non-minimally coupled to matter. We introduced a power-law term to the standard $f(Q)=Q$ theory such that in our suggested model we considered the functional form $\alpha Q^m$ and a linear in $Q$ non-minimal coupling. We analyzed the conditions for an expanding single-component universe which led to the constraint that the parameter $m$ must be greater than $1$.  We assumed baryonic matter, radiation and a family of Chaplygin gas models: i) Generalized Chaplygin Gas (GCG, characterized by $A_{gcg}$ and $n_{gcg}$), ii) Modified Chaplygin Gas (MCG, characterized by $A_{mcg}$, $n_{mcg}$ and $B$), iii) Variable Chaplygin Gas (VCG, characterized by $A_{vcg}$ and $n_{vcg}$), as constituents of the background fluid. We constrain these parameters using Markov Chain Monte Carlo (MCMC), leveraging the Quasar Dataset (QSO), Baryon-Acoustic Oscillations (BAO) measurements, and Observational Hubble Data (OHD).

We first calibrate the three QSO datasets from the years 2015, 2019, 2020 using an analytic expression for the error in comoving distance and obtain the constraints on these datasets using model-independent B\'ezier-style equation for Hubble parameter, as first introduced in \cite{Wei_2020}. Then using the 2019 calibrated values, we obtain the constraints on our proposed models. For all three models, we found $m>1$, affirming an expanding universe. The present value of baryonic matter density $\Omega_{b,0}$ remains consistent across models within $1-\sigma$. Except the GCG model, we found that MCG and VCG models, within the framework of non-minimal $f(Q)$ coupling, deviate from the standard $\Lambda$CDM strongly. We also used the AIC and BIC methods to statistically compare the performance of the three models with $\Lambda$CDM. For the GCG model, we found a moderate level of support with the AIC method, whereas the other two models (MCG and VCG) are rejected in favor of $\Lambda$CDM with AIC. With the BIC method, we find that $\Lambda$CDM is preferred over all three models, which aligns with expectations since BIC imposes a more stringent penalty on model complexity.

Through analysis of the deceleration parameter $q$, there is evidence for an accelerated expansion of the universe in all three models, in accordance with numerous studies such as \cite{riess1998observational,perlmutter1999measurements,percival2010baryon}. The transition redshift $z_t$ at which the universe switches from a decelerated to an accelerated phase was obtained for GCG, MCG and VCG models as $0.620^{+0.018}_{-0.017}$, $0.537^{+0.017}_{-0.017}$ and $0.470^{+0.012}_{-0.012}$ respectively. We also studied the jerk parameter $j$ to validate our model. The current jerk $j_0$ of our proposed models deviate from the $\Lambda$CDM universe at the current epoch. Analysis of the evolution of the density parameter highlights high MCG density at early times due to $B>0$, while the VCG model exhibits the strongest phantom-like behavior under the Om diagnostic at $z\le 1$.

In future work, one can also investigate other functional forms of $f(Q)$ as explored in several studies such as \cite{shabani2025emergent,Ghosh2024} who have discussed whether their proposed forms have stable solutions or not; as well as studies such as \cite{YADAV2024114,SULTANA2025100422} who have used cosmological datasets such as Pantheon+ to find parameters such as the deceleration parameter and transitions redshifts using their proposed functional forms. Finally, it should be mentioned that the results of our studies could be improved if more measurements of cosmic chronometers became available. Since there are only 31 cosmic-chronometer measurements available up to $z=2$, we predict a more accurate fit would play a significant role in making the calibrations more precise and setting tighter constraints on the parameters. 
Furthermore, we know that $f(Q)$ cosmology does not suffer from the cosmological constant problem and explains the current observations well. This sets the motivation to test various other dark energy candidates in the light of non-minimal coupling. This can be a dedicated topic for future research.  
\section{Acknowledgments}
\label{sec:ack}
We would like to thank Guido Risaliti who kindly allowed us access to the data set he and his collaborators compiled in 2019 (we refer to this in Sec. \ref{sec:quasardata} as the 2019 dataset).

F. S is grateful to the University of Tehran for supporting this work under a grant provided by the university research council.
N. A. and A. P. acknowledge that this work was carried out with the support of the graduate research assistant fellowship (GRAF) of the University of Alberta. 

This research has made use of the VizieR catalog access tool, CDS,
 Strasbourg, France (DOI : 10.26093/cds/vizier). The original description of the VizieR service was published in 2000, A\&AS 143, 23.

\appendix
\section{Analytic expressions for \texorpdfstring{$D_C$}{DC} and \texorpdfstring{$\sigma_{D_C}$}{sigmaDC}}
\label{app:dcal}
We begin by rewriting Eq. \ref{eq:hubble-nogeo} in the following way:
\begin{equation}
    H(z)= a\left(\frac{z}{z_m}\right)^2+b\left(\frac{z}{z_m}\right)+e,
\end{equation}
where $a= \beta_0- 2\beta_1+ \beta_2$, $b = -2\beta_0+2\beta_1$, and
$e = \beta_0$. According to the fit obtained in \cite{Amati_2019}, $\beta_0=H_0=67.76\pm 3.68$, $\beta_1=103.33\pm 11.16$, and $\beta_2=208.45\pm 14.29$, all in units of $\mathrm{km} \mathrm{s}^{-1} \mathrm{Mpc}^{-1}$. Since $b^2-4ae<0$, we define here $\Delta=\sqrt{4ae-b^2}$.
Substituting this Hubble expression in Eqn. \ref{eq:comoving_distance} and solving the integral, we get: 
\begin{equation}
\label{eqn:dc_analytic}
D_C(z)=\frac{2cz_m}{\Delta}\left(\tan^{-1}\left(\frac{\zeta(z)}{\Delta
}\right)-\tan^{-1}\left(\frac{b}{\Delta}\right)\right)
\end{equation}
$\zeta(z)=2a(z/z_m)+b$. In order to find the analytic expression for the error in $D_C$ we employ error propagation for Eqn. (\ref{eqn:dc_analytic}). Assuming no error in $z$, since we don't have access to their errors, we obtain the following expression for the error $\sigma_{D_C}$ after simplification
\begin{equation}
    \begin{split}
        \sigma^2_{D_c}(z)&=\left(-2\beta_2\frac{D_C(z)}{\Delta^2}+\frac{2cz_m}{\Delta}\left(\frac{2/\Delta+2b\beta_2/\Delta^3}{1+b^2/\Delta^2}+\frac{2((z/z_m)-1)/\Delta-2\beta_2\zeta(z)/\Delta^3}{1+\zeta^2(z)/\Delta^2}\right)\right)^2\sigma^2_{\beta_0}\\
        &+\left(4\beta_1\frac{D_C(z)}{\Delta^2}+\frac{2cz_m}{\Delta}\left(\frac{-2/\Delta-4b\beta_1/\Delta^3}{1+b^2/\Delta^2}+\frac{2(1-2(z/z_m))/\Delta+4\beta_1\zeta(z)/\Delta^3}{1+\zeta^2(z)/\Delta^2}\right)\right)^2\sigma^2_{\beta_1}\\
        &+\left(-2\beta_0\frac{D_C(z)}{\Delta^2}+\frac{2cz_m}{\Delta}\left(\frac{2b\beta_0/\Delta^3}{1+b^2/\Delta^2}+\frac{2(z/z_m)/\Delta-2\beta_0\zeta(z)/\Delta^3}{1+\zeta^2(z)/\Delta^2}\right)\right)^2\sigma^2_{\beta_2}
    \end{split}
\end{equation}
where $\sigma_{\beta_0}$, $\sigma_{\beta_1}$ and $\sigma_{\beta_2}$ are the errors in $\beta_0$, $\beta_1$ and $\beta_2$ respectively. 

\section{Comparison of different QSO Datasets}
\label{app:comparison_qso}
\begin{figure}[ht!]
	\centering                            
	\includegraphics[width=\linewidth]{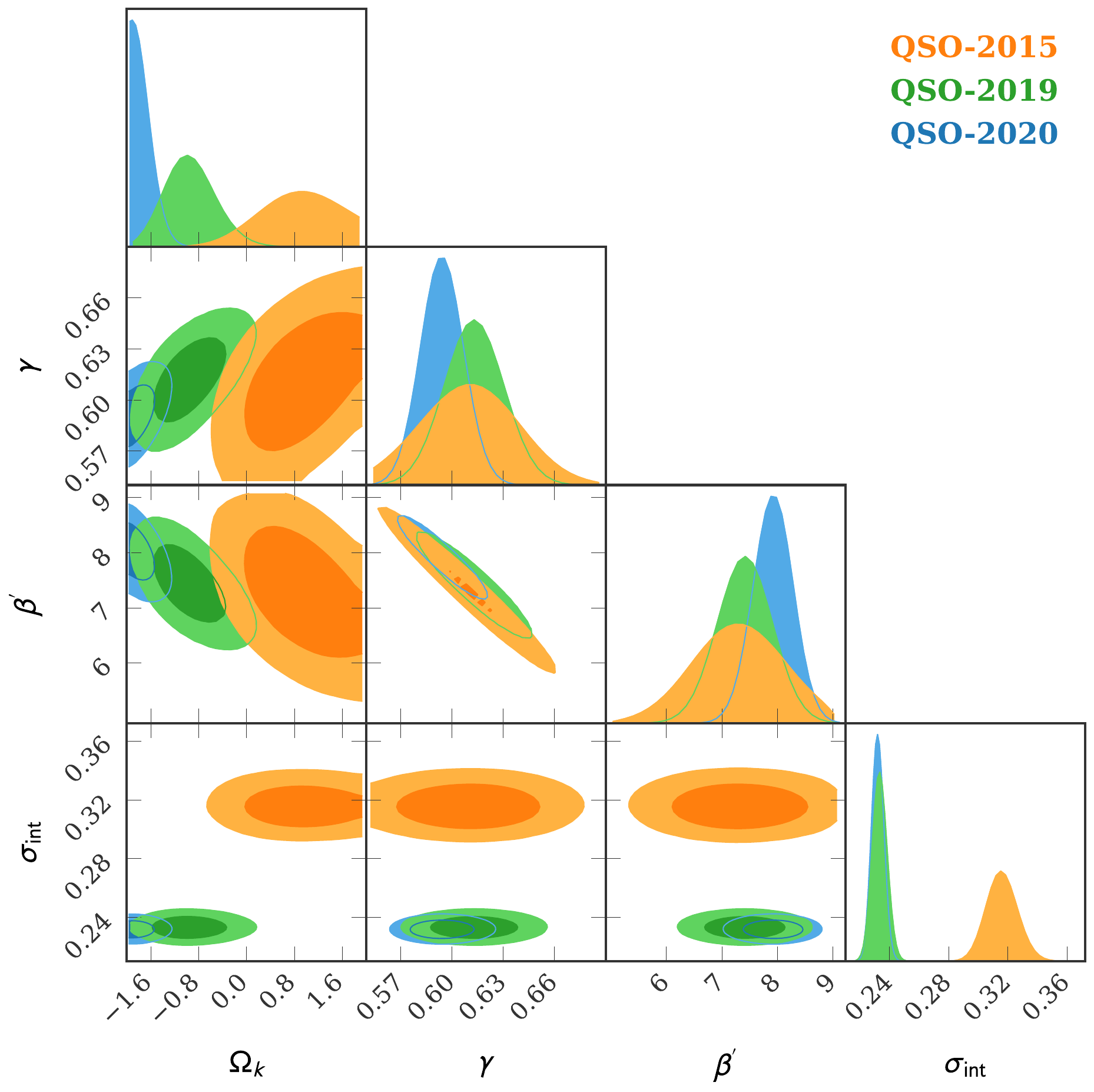}
	\caption{The corner plot above is comparing the values of $\{\Omega_k,\gamma,\beta^{'},\sigma_{\text{int}}\}$ for three different versions of the QSO datasets (each are color-coded according to the year the dataset was compiled, the years are seen in the legend). Each of the datasets were constrained by carrying out simulations with emcee.}
		\label{fig:3quasars}
\end{figure}
\bgroup
\def\arraystretch{1.5}
\begin{table}[ht!]
    \centering
    \begin{tabular}{ccccc}
    \hline
    QSO Dataset & $\Omega_k$ & $\gamma$ & $\beta'$ & $\sigma_{int}$\\
    \hline 
    2015 & $0.900^{+0.642}_{-0.675}$ &$0.611^{+0.028}_{-0.028}$ & $7.297^{+0.807}_{-0.807}$& $0.316^{+0.010}_{-0.010}$\\ 
    2019    & $-0.958^{+0.424}_{-0.383}$ &$0.613^{+0.017}_{-0.017}$ & $7.414^{+0.471}_{-0.474}$& $0.233^{+0.005}_{-0.005}$\\
    2020    & $-1.785^{+0.211}_{-0.146}$ &$0.594^{+0.012}_{-0.012}$ & $7.922^{+0.342}_{-0.347}$& $0.232^{+0.004}_{-0.004}$\\
    \hline
    \end{tabular}
    \caption{Comparison of the values of the 4 parameters that characterize the UV-X-ray flux relation for quasars, as explained in the main text, for each of the three QSO datasets.}
    \label{tab:circ3years}
\end{table}
In this section, we compare the calibration technique introduced in the main text for the three QSO measurements. The values of the four quasar parameters are highlighted in Table \ref{tab:circ3years}. One key observation about intrinsic dispersion is that in model-independent fitting, $\sigma_{int}$ decreases substantially between the QSO datasets of 2015 and 2019. Its value in the 2020 dataset is within $1\sigma$ of its 2019 counterpart fit. This is also very clear as seen from the corner plot in Fig. \ref{fig:3quasars}. In contrast, $\beta^{'}$ shows stability across all three datasets, with values that consistently fall within $1\sigma$ of each other. This robustness indicates that $\beta^{'}$ is largely unaffected by the increasing sample size of QSO. In the 2020 QSO dataset, the slope $\gamma$ shows a slight decrease compared to earlier datasets, again remaining within $1\sigma$ of the values from 2015 and 2019. However, the curvature parameter exhibits a much more pronounced change across all datasets. With an increasing number of quasar measurements, the curvature also becomes progressively negative. While the 2019 and 2020 measurements are consistent within 
$\sim2\sigma$, the 2020 value notably is very far from $\Omega_k=0$, signaling a significant departure from the typical expectation of flat cosmology. Due to this reason, we use the 2019 values of the calibrated parameters $\{\sigma_{int},\beta^{'},\gamma\}$ and set 
$\Omega_k=0$ to constrain the proposed $f(Q)$ model (as its result for $\Omega_k$ lies within a $3\sigma$ confidence from flat cosmology). 

\section{Analytic Expressions for Diagnostics}
\label{app:diagnostics}
\subsection{GCG and MCG}
In this section, we write the functional forms of $\{\phi_1(z), \phi_2(z), \phi_3(z)\}$ for the MCG model. The GCG equations can be obtained by setting $B=0$, $\Omega_{mcg,0}=\Omega_{gcg,0}$, $A_{mcg}=A_{gcg}$ and $n_{mcg}=n_{gcg}$. \\
Let $\Phi(z)=\left(A_{mcg}+(1-A_{mcg})(1+z)^{3(1+B)(1+n_{mcg})}\right)^{\frac{1}{1+n_{mcg}}}$. Therefore, the corresponding expressions for MCG in terms of $\Phi(z)$ are
\begin{equation}
    \begin{split}
\phi_1(z) &= \Omega_{b,0}(1+z)^3+\Omega_{r,0}(1+z)^4+\Omega_{mcg,0}\Phi(z)\\     
\phi_2(z) &=\Omega_{b,0}+\frac{4}{3}\Omega_{r,0}(1+z)+\Omega_{mcg,0}(1-A_{mcg})(1+B)(1+z)^{3\left(B+n_{mcg}(1+B)\right)}\Phi^{-n_{mcg}}(z)\\ 
\phi_3(z) &=\frac{4}{3}\Omega_{r,0}+3\Omega_{mcg,0}(1-A_{mcg})(B+n_{mcg}(1+B))(1+B)(1+z)^{3(B+n_{mcg}+Bn_{mcg})-1}\Phi^{-n_{mcg}}(z)\\
&\hspace{14pt}-3\Omega_{mcg,0}n_{mcg}(1-A_{mcg})^2(1+B)^2(1 + z)^ {2\bigl(3\left(B+n_{mcg}+Bn_{mcg}\right)+1\bigr)}\Phi^{-(2n_{mcg}+1)}(z)
    \end{split}
\end{equation}
\subsection{VCG}
In case of VCG, we redefine $\Phi(z)=\bigl(A_{vcg}(1+z)^6+(1-A_{vcg})(1+z)^{n_{vcg}}\bigr)^{\frac{1}{2}}$ such that
\begin{equation}
    \begin{split}
\phi_1(z) &= \Omega_{b,0}(1+z)^3+\Omega_{r,0}(1+z)^4+\Omega_{vcg,0}\Phi(z)\\     
\phi_2(z) &=\Omega_{b,0}+\frac{4}{3}\Omega_{r,0}(1+z)+\Omega_{vcg,0}\Bigl(A_{vcg}(1+ z)^3 + \frac{1}{6}n_{vcg}(1 - A_{vcg})(1 + z)^ {n_{vcg} - 3}\Bigr)
\Phi^{-1}(z)\\ 
\phi_3(z) &=\frac{4}{3}\Omega_{r,0}+\Omega_{vcg,0}\Bigl(3A_{vcg}(1+ z)^2 + \frac{1}{6}n_{vcg}(n_{vcg}-3)(1 - A_{vcg})(1 + z)^ {n_{vcg} - 4}\Bigr)\Phi^{-1}(z)\\
&\hspace{14pt}-\Omega_{vcg,0}\tau(z)\Bigl(A_{vcg}(1+ z)^3 + \frac{1}{6}n_{vcg}(1 - A_{vcg})(1 + z)^ {n_{vcg} - 3}\Bigr)\Phi^{-3}(z)
    \end{split}
\end{equation}
where $\tau(z)=3A_{vcg}(1+z)^5+(n_{vcg}/2)((1-A_{vcg})(1+z)^{n_{vcg}-1})$.

\bibliographystyle{elsarticle-num-names} 

\bibliography{refs}

\end{document}